\documentclass[aip,jmp,reprint,a4paper,amsmath,onecolumn]{revtex4-2}
\usepackage[utf8]{inputenc}
\usepackage{amssymb}
\usepackage{enumitem}
\usepackage{mathrsfs}
\usepackage{multirow} 
\usepackage{stackengine}
\usepackage[colorlinks=true,urlcolor=blue,anchorcolor=blue,citecolor=blue,filecolor=blue,linkcolor=blue,menucolor=blue,linktocpage=true,pdfa=true,unicode]{hyperref}
\usepackage{bbm}
\usepackage{bm}
\usepackage{lmodern}
\usepackage{graphicx}
\usepackage{ulem}

\newcommand\xrowht[2][0]{\addstackgap[.5\dimexpr#2\relax]{\vphantom{#1}}}

\newcommand{\DD}{\dfrac{d^2}{d\rho^2}}

\newcommand{\dom}[1]{\mathrm{Dom}(#1)}

\newcommand{\inpr}[2]{\langle #1,#2\rangle}
\newcommand{\comment}[1]{}

\renewcommand{\emph}[1]{{\it #1}}

\begin{document}
\title{Scalar field in $\mathrm{AdS}_2$ and representations of $\widetilde{\mathrm{SL}}(2,\mathbb{R})$}

\author{Atsushi Higuchi}
\email{atsushi.higuchi@york.ac.uk}
\affiliation{Department of Mathematics, University of York, Heslington, York, YO10 5DD, United Kingdom}
\author{Lasse Schmieding}
\email{lcs549@york.ac.uk}
\affiliation{Department of Mathematics, University of York, Heslington, York, YO10 5DD, United Kingdom}
\author{David Serrano Blanco}
\email{dsb523@york.ac.uk}
\affiliation{Department of Mathematics, University of York, Heslington, York, YO10 5DD, United Kingdom}

\date{November 2022}

\begin{abstract}
We study the solutions to the Klein-Gordon equation for the massive scalar field in the universal covering space of two-dimensional anti-de Sitter space. For certain values of the mass parameter, we impose a suitable set of boundary conditions which make the spatial component of the Klein-Gordon operator self-adjoint. This makes the time-evolution of the classical field well defined. Then, we use the  transformation properties of the scalar field under the isometry group of the theory, namely, the universal covering group of $\mathrm{SL}(2,\mathbb{R})$, in order to determine which self-adjoint boundary conditions are invariant under
this group, and which lead to the positive-frequency solutions forming a unitary representation of this group and, hence, to a vacuum state invariant under this group. Then we examine the cases where the boundary condition leads to an invariant theory with non-invariant vacuum state and determine the unitary representation to which the vacuum state belongs.
\end{abstract}

\maketitle

\section{Introduction}\label{Intro}

Anti-de Sitter space plays an important role in many areas of theoretical physics, most prominent among them being the AdS/CFT correspondence.\cite{maldacena} This has led to investigation of its properties and applications in a wide variety of contexts. One of the peculiarities of anti-de Sitter space is that it is not a globally hyperbolic manifold. Nevertheless, as clarified by Ishibashi and Wald,~\cite{ishi1,ishi2,ishi3} it is possible to define sensible and deterministic field dynamics when appropriate boundary conditions are imposed on the fields. They analyzed the scalar, vector and symmetric tensor field theories defined on $n$-dimensional anti-de Sitter space ($\mathrm{AdS}_n$), or more precisely, its universal covering space, for $n\geq 3$. (We shall refer to the universal covering space of anti-de Sitter space as anti-de Sitter space, $\mathrm{AdS}_n$, from now on.) A family of boundary conditions was found by applying the theory of self-adjoint extensions to the  radial component of the Klein-Gordon operator, resulting in a family of theories parametrized by a real number. Some properties of field theories with these boundary conditions have been studied recently.\cite{dappiaggi,morley}

In this paper we study free scalar field theory in two-dimensional anti-de Sitter space, $\mathrm{AdS}_2$. In this spacetime there are two disjoint spatial boundaries unlike in the higher-dimensional case studied by Ishibashi and Wald. Due to this fact the self-adjoint extensions for the spatial Klein-Gordon operator in $\mathrm{AdS}_2$ are richer than in the higher-dimensional case.  The results of Ishibashi and Wald give some special self-adjoint extensions in two dimensions but not all of them. This is because, for low values of the squared mass, the self-adjoint extensions in $\mathrm{AdS}_n$ with $n \geq 3$ are parametrized by one real number whereas the extensions for the two-dimensional case are parametrized by a $2\times 2$ unitary matrix.

We also note that not all of the consistent theories may be of physical interest.  Depending on the context in which such theories are analyzed, different arguments may be given for choosing a particular theory over the others. 
One possibility is to require the invariance under the isometry group of the spacetime. Applying this requirement to scalar field theory in $\mathrm{AdS}_2$, among the family of different consistent theories arising from self-adjoint boundary conditions we may choose those whose positive-frequency solutions form a unitary irreducible representation 
of the symmetry group of $\mathrm{AdS}_2$, \textit{i.e.}\ the universal covering group of $\mathrm{SL}(2,\mathbb{R})$ denoted by $\widetilde{\mathrm{SL}}(2,\mathbb{R})$.  
All unitary irreducible representations (UIRs) of $\widetilde{\mathrm{SL}}(2,\mathbb{R})$ have been obtained by Pukanzki~\cite{puk} using an analysis similar to that of Bargmann~\cite{barg} for the case of $\mathrm{SL}(2,\mathbb{R})$ and are well known.\cite{lang,knapp,har,rep,Kitaev}  In this paper, we identify the self-adjoint boundary conditions that are invariant under $\widetilde{\mathrm{SL}}(2,\mathbb{R})$.  Then, we identify the boundary conditions among these that lead to an $\widetilde{\mathrm{SL}}(2,\mathbb{R})$-invariant positive-frequency subspace, which corresponds to an $\widetilde{\mathrm{SL}}(2,\mathbb{R})$-invariant vacuum state. We also study the cases where the boundary conditions are invariant but the vacuum state is not and identify the UIR to which the vacuum state belongs.

Free scalar and spinor field theories have previously been studied by Sakai and Tanii.\cite{sak}
In their analysis the boundary conditions for the mode functions are determined by imposing 
the vanishing of energy flux at the conformal boundaries. As we shall show, the boundary conditions stemming from the energy flux condition coincide with the  self-adjoint extensions corresponding to the boundary conditions invariant under the action of $\widetilde{\mathrm{SL}}(2,\mathbb{R})$. 

Thus, we analyze a scalar field of mass $M$ obeying the Klein-Gordon equation in $\mathrm{AdS}_2$.  More specifically, we study the self-adjoint boundary conditions for this equation and find those such that the positive-frequency solutions form a UIR of $\widetilde{\mathrm{SL}}(2,\mathbb{R})$. The type of self-adjoint boundary conditions depends on the value of the mass of the field. If the mass is sufficiently large, the boundary conditions are uniquely determined by requiring the solutions to the Klein-Gordon equation to be normalizable with respect to the Klein-Gordon inner product, while in a certain range of low mass parameter the boundary conditions need to be specified.
The theory of self-adjoint extensions will be used to obtain such boundary conditions. Then we determine for which self-adjoint extensions the corresponding positive-frequency solutions form UIRs. 
It will be found that only the generalized Dirichlet and Neumann boundary conditions preserve the symmetry of anti-de Sitter space for the positive-frequency subspace of the solution space.

The rest of the paper is organized as follows. In Sec.~\ref{CADS} we describe the geometry of $\mathrm{AdS}_2$ and give our notations and conventions. In Sec.~\ref{UIRs} we briefly discuss the classification of the UIRs of $\widetilde{\mathrm{SL}}(2,\mathbb{R})$, using the viewpoint and notation of Kitaev.\cite{Kitaev} In Sec.~\ref{Scalar} we begin the analysis of the solutions to the Klein-Gordon equation. We apply the theory of self-adjoint extensions due to Weyl~\cite{Weyl} and von~Neumann~\cite{Neu} to the spatial Klein-Gordon differential operator in order to obtain the self-adjoint boundary conditions. Then, we determine the boundary conditions which respect the $\widetilde{\mathrm{SL}}(2,\mathbb{R})$ symmetry and identify the UIRs the positive-frequency solutions may form when the boundary conditions respect the symmetry. In Sec.~\ref{sec-bogoliubov} we examine the cases where the boundary condition is invariant but the vacuum state is not and determine the UIR to which the vacuum state belongs.  We summarize our results in Sec.~\ref{sec-conclusion}.
In Appendix~\ref{App-static-spacetime} we review free quantum scalar field theory with a stationary vacuum state in general static spacetime. In Appendix~\ref{App-negative-definite} we show that the radial component of the Klein-Gordon operator is unbounded from below if the mass $M^2$ is less than $-1/4$.
 In Appendix~\ref{App-A-bar-boundary} we present some technical details about the closure of the spatial Klein-Gordon operator. 
 In Appendix~\ref{appendix} we discuss the relation between the two descriptions of self-adjoint extensions of the spatial Klein-Gordon operator, one in terms of eigenfunctions with imaginary eigenvalues and the other in terms of boundary conditions.
 In Appendix~\ref{App-negative-eigenvalues} we present a simple example of boundary conditions which lead to negative eigenvalues of the spatial Klein-Gordon operator.

\section{Two-dimensional anti-de Sitter space}\label{CADS}
Two-dimensional anti-de Sitter space is the hyperboloid embedded in three-dimensional flat space with two timelike coordinates $x^0$, $x^1$ and one spacelike coordinate $x^2$ given by the equation
\begin{align}
\left(x^0\right)^2+\left(x^1\right)^2-\left(x^2\right)^2=1\,.
\end{align}
We choose the local coordinates, $\rho\in(-\pi/2,\pi/2)$ and $t\in(-\pi,\pi]$, as
\begin{subequations}\label{coords}
\begin{align}
x^0&=\sec\rho\cos t\,,\\
x^1&=\sec\rho\sin t\,,\\
x^2&=\tan\rho\,.
\end{align}
\end{subequations}
The metric for this spacetime in coordinates given by Eq.~\eqref{coords} is
\begin{align}\label{metric}
d s^2=g_{ij}d x^{i}d x^{j}=\sec^2 \rho\left(-d t^2+d\rho^2\right)\,,
\end{align}
which is conformally flat. By comparing this metric with the general static metric~\eqref{AppA-static}, we have $N=\sqrt{g_{\rho\rho}}=\sqrt{g} = \sec\rho$. The universal covering space of anti-de Sitter space is the spacetime with this metric with  $t\in \mathbb{R}$.  This universal covering space~\cite{kroon} is denoted by $\mathrm{AdS}_2$ throughout this paper. The metric~\eqref{metric} makes it clear that the light-ray from the conformal boundary $\rho=\pi/2$ travels to $\rho=0$ in a finite coordinate time $\Delta t = \pi/2$.  Thus, indeed this spacetime is not globally hyperbolic.

The Killing vector fields in these coordinates are linear combinations of
\begin{subequations}\label{KillingVF}
\begin{align}
\Lambda_0&=\partial_t\,,\label{KillingVF-1}\\
\Lambda_1&=\cos\rho\sin t\,\partial_\rho+\sin\rho\cos t\,\partial_t\,,\label{KillingVF-2}\\
\Lambda_2&=\cos\rho\cos t\,\partial_\rho-\sin\rho\sin t\,\partial_t\,.\label{KillingVF-3}
\end{align}
\end{subequations}
These Killing vectors satisfy the $\mathfrak{sl}(2,\mathbb{R})$ commutation relations:
\begin{align}\label{Kcommutation}
[\Lambda_0,\Lambda_1]=\Lambda_2\,,\hspace{.5cm}[\Lambda_0,\Lambda_2]=-\Lambda_1\,,\hspace{.5cm}[\Lambda_1,\Lambda_2]=-\Lambda_0\,.
\end{align}

The Laplace-Beltrami operator for the metric~\eqref{metric} reads
\begin{align}
\Box&= -\frac{1}{N^2}\partial_t^2 + \frac{1}{\sqrt{g}\,N}\partial_\rho \sqrt{g}\,Ng^{\rho\rho}\partial_\rho\,,\nonumber \\
&=\cos^2\rho\left(-\partial_{t}^{2}+\partial_{\rho}^{2}\right)\,,
\end{align}
and it is easy to verify that the quadratic Casimir operator of $\mathfrak{sl}(2,\mathbb{R})$ is
\begin{align}
Q&:=-\Lambda_0^2+\Lambda_1^2+\Lambda_2^2\,,\nonumber\\
&=\Box\,.\label{Casimir1}
\end{align}

\section{Unitary irreducible representations of $\widetilde{\mathrm{SL}}(2,\mathbb{R})$}\label{UIRs}

In this section we identify the isometry group of $\mathrm{AdS}_2$ via the algebra generated by the Killing vector fields given by Eq.~\eqref{KillingVF}. We adopt a notation similar to that of Kitaev.\cite{Kitaev}
The raising and lowering operators, $L_+$ and $L_{-}$, and the operator $L_0$ are defined by
\begin{subequations}\label{ladder}
\begin{align}
    L_\pm & = \Lambda_1 \pm i\Lambda_2\,,\\
    L_0 & = i \Lambda_0\,.
\end{align}
\end{subequations}
They satisfy the following commutation relations:
\begin{align}\label{Lcommut}
[L_0,L_{+}]=L_{+}\,,\hspace{.5cm}[L_0,L_{-}]=-L_{-}\,,\hspace{.5cm}[L_{+},L_{-}]=2L_0\,.
\end{align}
The group $\widetilde{\mathrm{SL}}(2,\mathbb{R})$ is the unique simply connected group obtained by the exponential map of this Lie algebra.  We note, in particular, that the one-dimensional subgroup generated by $L_0$ has the topology of $\mathbb{R}$ rather than a circle.
The quadratic Casimir operator of $\mathfrak{sl}(2,\mathbb{R})$ in the basis given by Eq.~\eqref{ladder} is 
\begin{align}\label{LCasimir}
Q=L_0^2+\dfrac{1}{2}\left(L_+L_- +L_-L_+\right)\,.
\end{align}

Now we describe the UIRs of $\widetilde{\mathrm{SL}}(2,\mathbb{R})$. By Schur's lemma, the quadratic Casimir operator is constant over an irreducible representation.\cite{knapp}  Let $q:=\lambda(\lambda-1)$ be the eigenvalue of the Casimir operator $Q$ in a given representation. If this representation is unitary, then the operators $\Lambda_i$, $i=0,1,2$, are skew-symmetric, \textit{i.e.}\ anti-Hermitian, with respect to the inner product and, hence, the eigenvalue $q$ is real. This allows us to restrict the values of $\lambda$ as
\begin{align}\label{lambda-range}
\lambda\in\dfrac{1}{2}+i\mathbb{R}^{+}\,\hspace{.5cm}\text{or}\hspace{.5cm}\lambda\in\mathbb{R}\  \textrm{and}\ \lambda \geq \frac{1}{2}\,,
\end{align}
for which we have $q<-1/4$ and $q\geq -1/4$, respectively.  We have taken into account the fact that $q$ is invariant under the transformation $\lambda \mapsto 1-\lambda$ in restricting the values for $\lambda$ in Eq.~\eqref{lambda-range}.

The group $\widetilde{\mathrm{SL}}(2,\mathbb{R})$ has a center $Z=\{\exp(2\pi n \Lambda_0)|n\in\mathbb{Z}\}$ generated by the element $\exp(2\pi \Lambda_0)$.  The eigenvalue $e^{-2\pi i \mu}$ of this generator is also constant over a UIR by Schur's lemma.  Let $\phi_\omega$ be an eigenvector of $L_0 = i\Lambda_0$ with eigenvalue $\omega$ in a UIR with the eigenvalue of $\exp(2\pi \Lambda_0)$ being $e^{-2\pi i \mu}$.  Then, $\omega = \mu + k$ with $k\in \mathbb{Z}$.  
Now, one finds from the commutation relations~\eqref{Lcommut} that $L_0L_\pm \phi_\omega = (\omega\pm 1)L_\pm\phi_\omega$.  We also find from Eqs.~\eqref{Lcommut} and \eqref{LCasimir} that
\begin{subequations}\label{Lmatrix}
\begin{align}
\inpr{L_{-}\phi_\omega}{L_{-}\phi_\omega} & = -\inpr{\phi_\omega}{L_+L_-\phi_\omega}=-q+\omega^2-\omega\,,\label{Lmatrix-1}\\
\inpr{L_{+}\phi_\omega}{L_{+}\phi_\omega} & = - \inpr{\phi_\omega}{L_-L_+\phi_\omega}=-q+\omega^2+\omega\,,\label{Lmatrix-2}
\end{align}
\end{subequations}
because $L_{\pm}^\dagger = - L_{\mp}$ by Eq.~\eqref{ladder}.
If the representation is unitary, then the right-hand side of Eq.~\eqref{Lmatrix} must be non-negative. Recalling $\omega=\mu+k$, $k\in\mathbb{Z}$ and $q=\lambda(\lambda-1)$, we can write this requirement as
\begin{subequations}\label{muineq}
\begin{align}
(k+\mu-\lambda)(k+\mu+\lambda-1)&\geq 0\,,\label{muineq-1}\\
(k+\mu+\lambda)(k+\mu-\lambda+1)&\geq 0\,,\label{muineq-2}
\end{align} 
\end{subequations}
for the value of $k$ for every eigenvector $\phi_\omega$ in the given representation.  Notice that Eq.~\eqref{muineq-2} is obtained from Eq.~\eqref{muineq-1} by letting $(k,\mu) \mapsto (-k,-\mu)$.

The non-trivial UIRs are labeled by $(\lambda,\mu)$.  Some of them are found by requiring that Eqs.~\eqref{muineq} are satisfied by all $k\in\mathbb{Z}$ whereas others are found by requiring that both Eqs.~\eqref{muineq-1} and \eqref{muineq-2} are satisfied for $k \in \mathbb{N}$ and that the equality in Eq.~\eqref{muineq-1} is satisfied by $k=0$, or that Eqs.~\eqref{muineq-1} and \eqref{muineq-2} are satisfied for $-k \in \mathbb{N}$ and that the equality in Eq.~\eqref{muineq-2} is satisfied by $k=0$. In this manner, one finds the following UIRs, which exhaust all non-trivial UIRs up to isomorphisms:
\begin{enumerate}
\item {\bf Discrete series representations:} $\mathscr{D}_{\lambda}^{\pm}$ for $\lambda>1/2$, with $\mu=\pm\lambda$, and $\omega=\pm(\lambda+k)$, respectively, where $k\in\mathbb{N}\cup\{0\}$; $\mathscr{D}_{1-\lambda}^{\pm}$ for $1/2<\lambda<1$, with $\mu=\pm(1-\lambda)$ and $\omega=\pm(1-\lambda+k)$, respectively, where $k\in\mathbb{N}\cup\{0\}$. The Casimir eigenvalue satisfies $q>-1/4$.
\item {\bf Principal series representations:} $\mathscr{P}_{is}^{\mu}$ for $\lambda=1/2+is$, with $s\in\mathbb{R}^+$, $- 1/2<\mu \leq 1/2$, and $\omega=\mu+k$, where $k\in\mathbb{Z}$. The Casimir eigenvalue satisfies $q<-1/4$.
\item {\bf Complementary series representations:} $\mathscr{C}_{\lambda}^{\mu}$ for $0<\lambda<1/2$, with , $|\mu|<\lambda$, and $\omega=\mu+k$, where $k\in\mathbb{Z}$. The Casimir eigenvalue satisfies $-1/4<q<0$.
\item {\bf Mock-discrete series representations:} $\mathscr{D}_{1/2}^{\pm}$ for $\mu=\lambda=1/2$, and $\omega=\pm(1/2+k)$, respectively, with $k\in\mathbb{N}\cup\{0\}.$ The Casimir eigenvalue is $q=-1/4$.
\end{enumerate}

\section{Scalar field theory in $\mathrm{AdS}_2$}\label{Scalar}

We now analyze a scalar field obeying the Klein-Gordon equation in the $\mathrm{AdS}_2$ background, \textit{i.e.}\  $(\Box-M^2)\phi(t,\rho)=0$, which in local coordinates given by Eq.~\eqref{coords} reads
\begin{align}\label{KG}
\cos^2\rho\left[-\frac{\partial^2}{\partial t^{2}}+\frac{\partial^{2}}{\partial\rho^{2}}-\frac{\lambda(\lambda-1)}{\cos^2\rho}\right]\phi(t,\rho)=0\,.
\end{align}
Here we have let $M^2=\lambda(\lambda-1)$ so that it is identified with the Casimir eigenvalue $q$ from Sec.~\ref{CADS}. Given two solutions, $\phi^{(1)}(t,\rho)$ and $\phi^{(2)}(t,\rho)$, to this equation, the following inner product, the Klein-Gordon inner product, is time-independent:
\begin{equation}\label{KG-original-def}
    \langle \phi^{(1)},\phi^{(2)}\rangle_{\mathrm{KG}} :=
    i\int_{-\pi/2}^{\pi/2}\left[ \overline{\phi^{(1)}(t,\rho)}\frac{\partial\phi^{(2)}(t,\rho) }{\partial t} - \frac{\partial\overline{\phi^{(1)}(t,\rho)}}{\partial t}\phi^{(2)}(t,\rho)\right]d\rho\,.
\end{equation}

We first describe the solutions to Eq.~\eqref{KG} of the form
\begin{align}\label{modeform}
\phi(t,\rho)=\Psi_\omega(\rho)e^{-i\omega t}\,,\hspace{.5cm}\omega > 0\,.
\end{align}
The spatial component satisfies
\begin{align}\label{spatialKG}
A\Psi_\omega(\rho)=\omega^2\Psi_\omega(\rho)\,,
\end{align}
where we have defined the differential operator
\begin{align}\label{OpA}
A:=- \DD+\dfrac{\lambda(\lambda-1)}{\cos^2\rho}\,.
\end{align}
Two independent solutions of Eq.~\eqref{spatialKG} are given in terms of the Gaussian hypergeometric functions~\cite{grad} and read
\begin{subequations}\label{Solutions}
\begin{align}
\Psi_\omega^{(1)}(\rho)&=(\cos\rho)^{\lambda}\,F\left(\frac{\lambda+\omega}{2},\frac{\lambda-\omega}{2};\frac{1}{2};\sin^{2}\rho\right)\,,\label{Solutions-1}\\
\Psi_\omega^{(2)}(\rho)&=\sin\rho\,(\cos\rho)^{\lambda}\,F\left(\frac{1+\lambda+\omega}{2},\frac{1+\lambda-\omega}{2};\frac{3}{2};\sin^{2}\rho\right)\,.\label{Solutions-2}
\end{align}
\end{subequations}
(Strictly speaking, the operator $A$ in Eq.~\eqref{spatialKG} should be $A^\dagger$.  See below.)

As shown in Appendix~\ref{App-static-spacetime}, the relevant inner product for these solutions is
\begin{align}\label{space-KG-inner}
\langle \Psi_1,\Psi_2\rangle := \int_{-\pi/2}^{\pi/2}\overline{\Psi_1(\rho)}\Psi_2(\rho)\,d\rho\,.
\end{align}
(Let $N=\sqrt{g}=\sec\rho$ in Eq.~\eqref{AppA-inner-product}.)
Thus, we are led to study the properties of the operator $A$ defined by Eq.~\eqref{OpA} with this inner product.
The operator $A$ in Eq.~\eqref{OpA} is symmetric, \textit{i.e.}
\begin{align}
    \langle \Psi_1,A\Psi_2\rangle = \langle A\Psi_1,\Psi_2\rangle\,,
\end{align}
on the domain~\cite{reed,ishi3} $\dom{A}=C_c^{\infty}(-\pi/2,\pi/2)$, \emph{i.e.}\ the set of compactly supported smooth functions with support away from the boundaries. We note that $\dom{A}$ is dense in $L^2[-\pi/2,\pi/2]$, \emph{i.e.}\ that any function in the latter can be approximated by functions in $\dom{A}$ ``arbitrarily well''.
The adjoint $A^\dagger$ of the operator $A$ is defined as follows.\cite{reed-one} If there is a function $\psi\in L^2[-\pi/2,\pi/2]$ such that $\langle\Psi,A\Phi\rangle = \langle \psi,\Phi\rangle$ for all $\Phi\in C_c^\infty(-\pi/2,\pi/2)$, then $\Psi \in \dom{A^\dagger}\subseteq L^2[-\pi/2,\pi/2]$ and $A^\dagger\Psi = \psi$.  The set $\dom{A^\dagger}$ is the domain of $A^\dagger$.  Thus, if $\Psi \in \dom{A^\dagger}$, then
\begin{align}
    \langle \Psi,A\Phi\rangle = \langle A^\dagger \Psi,\Phi\rangle\,,
\end{align}
for all $\Phi \in \dom{A} = C_c^\infty(-\pi/2,\pi/2)$.
It is known that, if $\Psi\in \dom{A^\dagger}$, then the derivative $\Psi'$ exists in $(-\pi/2,\pi/2)$ and is absolutely continuous.\cite{reed} An important consequence of this fact is that the operator $A^\dagger$ is the same differential operator as $A$ on $\Psi\in \dom{A^\dagger}$ except on a measure-zero set, where $\Psi$ may not be twice differentiable, and that, if $\Psi_1, \Psi_2 \in \dom{A^\dagger}$, then the following equality from integration by parts holds:
\begin{align}\label{boundary-for-A-dagger}
    \langle A^\dagger \Psi_1,\Psi_2\rangle - \langle \Psi_1,A^\dagger \Psi_2\rangle =
    \left[ \overline{\Psi_1(\rho)}\frac{d\Psi_2(\rho) }{d\rho} - \frac{d\overline{\Psi_1(\rho)}}{d\rho} \Psi_2(\rho)\right]_{\rho\to -\pi/2}^{\rho\to \pi/2}\,.
\end{align}
We note also that $A^\dagger \Psi = A\Psi$ if $\Psi \in \dom{A} = C^\infty_c(-\pi/2,\pi/2)$.

An operator $\Omega \in L^2[-\pi/2,\pi/2]$ is \textit{self-adjoint} if $\dom{\Omega} =\dom{\Omega^\dagger}$ and if it is symmetric on this domain, \textit{i.e.}
\begin{align}
    \langle\Psi_1,\Omega\Psi_2\rangle = \langle \Omega\Psi_1,\Psi_2\rangle\,,
\end{align}
for all $\Psi_1,\Psi_2\in \dom{\Omega}$.
The operator $A$ is not self-adjoint because $\dom{A} \neq \dom{A^\dagger}$, the latter being larger.\cite{reed,hall}  To define a quantum theory of this scalar field as described in Appendix~\ref{App-static-spacetime} we need to find a self-adjoint operator $A_U$ with its domain satisfying $\dom{A} \subseteq \dom{A_U} \subseteq \dom{A^\dagger}$, such that $A_U\Psi = A^\dagger \Psi$ if $\Psi\in \dom{A_U}$.  If this is the case, then the operator $A_U$ is said to be a \textit{self-adjoint extension} of $A$.  Given a self-adjoint operator $A_U$ with positive spectrum, one can define a quantum theory with a stationary vacuum state for this scalar field.

The operator $A$ is positive for $\lambda \in \mathbb{R}$ because
\begin{equation}
   A = \left( - \frac{d\ }{d\rho} + \lambda\tan\rho\right)\left( \frac{d\ }{d\rho} + \lambda\tan\rho\right) + \lambda^2\,.
\end{equation}
It is shown in Appendix~\ref{App-negative-definite} that the operator $A$ is unbounded from below if $M^2 = \lambda(\lambda-1) < -1/4$, \textit{i.e.} if $\lambda$ is imaginary.  (The method for the proof is similar to the higher-dimensional case.~\cite{ishi3})
For this reason, we assume that $M^2\geq -1/4$ and, hence, that  $\lambda \in\mathbb{R}$ and $\lambda \geq 1/2$ from now on.

To find self-adjoint extensions of the operator $A$ 
we need to analyze the behavior of the solutions~\eqref{Solutions} at the boundaries $\rho=\pm\pi/2$.
We first find  for which values of $\lambda$ we have square-integrable solutions to Eq.~\eqref{spatialKG} by examining
the asymptotic behavior of these solutions at the boundaries.
It turns out that it is convenient to analyze them in the following three cases separately:
\begin{enumerate}[label=({\it \roman*}\,)]
\item $\lambda> 3/2$ with $\lambda\neq k+1/2$ for any $k\in\mathbb{N}$;
\item $\lambda=1/2$ and $\lambda=k+1/2$ for $k\in\mathbb{N}$;
\item $1/2<\lambda< 3/2$.
\end{enumerate}
For the cases ({\it i}) and ({\it iii}), we use the following transformation formulas for the hypergeometric function~\cite{grad}, which are valid for $\lambda\neq k+\frac{1}{2}$, $k\in \mathbb{Z}$:
\begin{subequations}\label{expansion}
\begin{align}
\Psi_\omega^{(1)}(\rho)=&(\cos\rho)^{\lambda}\,A_{1}(\omega)F\left(\frac{\lambda+\omega}{2},\frac{\lambda-\omega}{2};\frac{1}{2}+\lambda;\cos^{2}\rho\right)\,,\nonumber\\
&+(\cos\rho)^{1-\lambda}\,B_{1}(\omega)\,F\left(\frac{1-\lambda+\omega}{2},\frac{1-\lambda-\omega}{2};\frac{3}{2}-\lambda;\cos^{2}\rho\right)\,,\label{expansion1}\\
\Psi_\omega^{(2)}(\rho)=&\sin\rho\,\left[(\cos\rho)^{\lambda}\,A_{2}(\omega)\,F\left(\frac{1+\lambda+\omega}{2},\frac{1+\lambda-\omega}{2};\frac{1}{2}+\lambda;\cos^{2}\rho\right)\,,\right.\nonumber\\
&+\left.(\cos\rho)^{1-\lambda}\,B_{2}(\omega)\,F\left(\frac{2-\lambda+\omega}{2},\frac{2-\lambda-\omega}{2};\frac{3}{2}-\lambda;\cos^{2}\rho\right)\right]\,,\label{expansion2}
\end{align}
\end{subequations}
where
\begin{align}\label{ABconstants}
A_{1}(\omega)&:=\frac{\Gamma\left(\frac{1}{2}\right)\Gamma\left(\frac{1}{2}-\lambda\right)}{\Gamma\left(\frac{1-\lambda+\omega}{2}\right)\Gamma\left(\frac{1-\lambda-\omega}{2}\right)}\,,\hspace{.5cm}B_1(\omega):=\frac{\Gamma\left(\frac{1}{2}\right)\Gamma\left(\lambda-\frac{1}{2}\right)}{\Gamma\left(\frac{\lambda+\omega}{2}\right)\Gamma\left(\frac{\lambda-\omega}{2}\right)}\,,\nonumber\\
A_2(\omega)&:=\frac{\Gamma\left(\frac{3}{2}\right)\Gamma\left(\frac{1}{2}-\lambda\right)}{\Gamma\left(\frac{2-\lambda+\omega}{2}\right)\Gamma\left(\frac{2-\lambda-\omega}{2}\right)}\,,\hspace{.5cm}B_2(\omega):=\frac{\Gamma\left(\frac{3}{2}\right)\Gamma\left(\lambda-\frac{1}{2}\right)}{\Gamma\left(\frac{1+\lambda+\omega}{2}\right)\Gamma\left(\frac{1+\lambda-\omega}{2}\right)}\,.
\end{align}
We define the variable $\widetilde{\rho}:=\pi/2-|\rho|$. Near the spatial boundaries, $\rho\to\pm\pi/2$, we have $\widetilde{\rho}\to 0$, and $\cos\rho\approx\widetilde{\rho}$. Since $F(a,b;c;x)=1+O(x)$ for $|x| \ll 1$, the behavior of the general solution $\Psi(\rho)=C_1\Psi_\omega^{(1)}(\rho)+C_2\Psi_\omega^{(2)}(\rho)$, with $C_1,C_2\in\mathbb{C}$, of Eq.~\eqref{spatialKG} for the cases ({\it i}) and ({\it iii}) is given by
\begin{align}\label{asymptotic}
\Psi(\rho)\sim \widetilde{\rho}^{\lambda}(C_1A_1(\omega)\pm C_2A_2(\omega)+O(\widetilde{\rho}^2))+\widetilde{\rho}^{1-\lambda}(C_1B_1(\omega)\pm C_2B_2(\omega)+O(\widetilde{\rho}^2))\,.
\end{align}

For the solutions with $\lambda=k+1/2$ in case ({\it ii}\,) the transformation formulas given by Eqs.~\eqref{expansion} and \eqref{ABconstants} are ill-defined. For this case the following transformation formulas are used instead~\cite{abr} 
\begin{subequations}
\begin{align}\label{transf1}
\Psi_\omega^{(1)}(\rho)=&H_1(\omega)(\cos\rho)^{k+\frac{1}{2}}\sum_{j=0}^{\infty}\frac{\left(\frac{k+\omega}{2}+\frac{1}{4}\right)_j\left(\frac{k-\omega}{2}+\frac{1}{4}\right)_j}{j!(j+k)!}(\cos\rho)^{2j}\left(\ln(\cos^2\rho)+h_1(j)\right)\,,\nonumber\\
&+B_1(\omega)(\cos\rho)^{-k+\frac{1}{2}}\sum_{j=0}^{k-1}\frac{\left(\frac{-k+\omega}{2}+\frac{1}{4}\right)_j\left(\frac{-k-\omega}{2}+\frac{1}{4}\right)_j}{j!(1-k)_j}(\cos\rho)^{2j}\,,
\end{align}
\begin{align}\label{transf2}
\Psi_\omega^{(2)}(\rho)=&H_2(\omega)\sin\rho(\cos\rho)^{k+\frac{1}{2}}\sum_{j=0}^{\infty}\frac{\left(\frac{k+\omega}{2}+\frac{3}{4}\right)_j\left(\frac{k-\omega}{2}+\frac{3}{4}\right)_j}{j!(j+k)!}(\cos\rho)^{2j}\left(\ln(\cos^2\rho)+h_2(j)\right)\,,\nonumber\\
&+B_2(\omega)\sin\rho(\cos\rho)^{-k+\frac{1}{2}}\sum_{j=0}^{k-1}\frac{\left(\frac{-k+\omega}{2}+\frac{3}{4}\right)_j\left(\frac{-k-\omega}{2}+\frac{3}{4}\right)_j}{j!(1-k)_j}(\cos\rho)^{2j}\,,
\end{align}
\end{subequations}
where we have defined
\begin{align}\label{Hconst}
H_1(\omega)&=\frac{(-1)^{k+1}\Gamma(1/2)}{\Gamma\left(\frac{-k+\omega}{2}+\frac{1}{4}\right)\Gamma\left(\frac{-k-\omega}{2}+\frac{1}{4}\right)}\,,\hspace{.5cm}
B_1(\omega)=\frac{\Gamma(k)\Gamma(1/2)}{\Gamma\left(\frac{k+\omega}{2}+\frac{1}{4}\right)\Gamma\left(\frac{k-\omega}{2}+\frac{1}{4}\right)}\,,\nonumber\\
H_2(\omega)&=\frac{(-1)^{k+1}\Gamma(3/2)}{\Gamma\left(\frac{-k+\omega}{2}+\frac{3}{4}\right)\Gamma\left(\frac{-k-\omega}{2}+\frac{3}{4}\right)}\,,\hspace{.55cm}
B_2(\omega)=\frac{\Gamma(k)\Gamma(3/2)}{\Gamma\left(\frac{k+\omega}{2}+\frac{3}{4}\right)\Gamma\left(\frac{k-\omega}{2}+\frac{3}{4}\right)}\,,
\end{align}
and the constants $h_1(j)$ and $h_2(j)$ are given by
\begin{subequations}
\begin{align}
h_1(j)&=\psi\left(\frac{k+\omega}{2}+\frac{1}{4}+j \right)+\psi\left(\frac{k-\omega}{2}+\frac{1}{4}+j \right)-\psi(j+1)-\psi(j+k+1)\,,\\
h_2(j)&=\psi\left(\frac{k+\omega}{2}+\frac{3}{4}+j \right)+\psi\left(\frac{k-\omega}{2}+\frac{3}{4}+j \right)-\psi(j+1)-\psi(j+k+1)\,,
\end{align}
\end{subequations}
and $\psi(x)$ is the digamma function.\cite{grad} Note first that the leading behavior of these functions is the same as in case ({\it i}) if $k \geq 1$ ($\lambda \geq 3/2)$. For $k=0$ ($\lambda=1/2$) the leading behavior for $\Psi(\rho)=C_1\Psi_\omega^{(1)}(\rho)+C_2\Psi_\omega^{(2)}(\rho)$ is found as
\begin{align}\label{asympt}
\Psi(\rho)\sim & \,\widetilde{\rho}^{\,\,\frac{1}{2}}\left[\ln(\widetilde{\rho}^2)(C_1H_1(\omega)\pm C_2H_2(\omega))+(C_1H_1(\omega)h_{1}(0)\pm C_2H_2(\omega)h_{2}(0))\right] \notag \\
&+O(\widetilde{\rho}^{\frac{3}{2}}\ln(\widetilde{\rho}^2))\,.
\end{align}

Using Eqs.~\eqref{asymptotic} and \eqref{asympt} we can determine when we have square-integrable solutions for each value of $\lambda$ in cases ({\it i}\,)--({\it iii}\,) as follows.

\begin{enumerate}[label=({\it \roman*}\,)]
\item For this case we have $2\lambda> 3$. Hence, the first term in Eq.~\eqref{asymptotic} is square integrable. However, because $2-2\lambda< -1$, the second term is not square integrable unless it vanishes. Hence, the solution $\Psi(\rho)$ is square integrable if and only if $C_1B_1(\omega)\pm C_2B_2(\omega)=0$. This can be achieved for both $\rho=\pm \pi/2$ if and only if $B_1(\omega)=0$ and $C_2=0$, or $B_2(\omega)=0$ and $C_1=0$. From Eq.~\eqref{ABconstants} we find that the conditions $B_1(\omega)=0$ and $C_2=0$ give the even solution $\Psi_\omega^{(1)}(\rho)$ with $\omega=\lambda+2\ell$, $\ell\in \mathbb{N}\cup\{0\}$ while the conditions $B_2(\omega)=0$ and $C_1=0$ give the odd solution $\Psi_\omega^{(2)}(\rho)$ with $\omega=\lambda+2\ell+1$, $\ell\in\mathbb{N}\cup \{0\}$. These two cases can be combined to give the positive-frequency functions as~\cite{grad}
\begin{align}\label{case1sol}
\phi_n^{\mathrm{I}}(t,\rho)=N_{n}^{\mathrm{I}}(\cos\rho)^{\lambda}\,P_{n}^{\,(\lambda-1/2,\lambda-1/2)}(\sin\rho)e^{-i\omega_n^{\mathrm{I}}t}\,,\ \ \omega^{\mathrm{I}}=\lambda+n\,.
\end{align}
where $P_{n}^{(a,b)}(x)$, $n\in \mathbb{N}\cup \{0\}$, are the Jacobi polynomials defined by
\begin{align}
    P_n^{(a,b)}(x):= \frac{\Gamma(n+a+1)}{n!\Gamma(a+1)}F\left(n+a+b+1,-n;a+1;\frac{1-x}{2}\right)\,,
\end{align}
and $N_{n}^{\mathrm{I}}$ are normalization constants such that the mode functions $\phi_n^{\mathrm{I}}(t,\rho)$ are normalized by the Klein-Gordon inner product~\eqref{KG-original-def},
\begin{align}
    \langle \phi_m^{\mathrm{I}},\phi_n^{\mathrm{I}}\rangle_{\mathrm{KG}} =
    2\omega_n\langle \Psi_{m}^{\mathrm{I}},\Psi_{n}^{\mathrm{I}}\rangle = \delta_{mn}\,,
\end{align}
if we write $\phi_{n}^{\mathrm{I}}(t,\rho)=\Psi_n^{\mathrm{I}}(\rho)e^{-i\omega_n^{\mathrm{I}} t}$.  These constants are found by using the standard normalization integral for the Jacobi polynomials (see e.g., Eq.~7.391 of Gradshteyn and Ryzhik~\cite{grad}),
\begin{align}\label{Jacobi-normalisation}
    & \int_{-1}^1 (1-x)^a(1+x)^b P_n^{(a,b)}(x)P_m^{(a,b)}(x)\,dx\notag \\
    & = \frac{2^{a+b+1}\Gamma(a+n+1)\Gamma(b+n+1)}{n!(a+b+1+2n)\Gamma(a+b+n+1)}\delta_{nm}\,,\ \ a, b > -1\,.
\end{align}
as
\begin{align}
    N_n^{\mathrm{I}} = \frac{\sqrt{n!\Gamma(2\lambda+n)}}{2^{\lambda}\Gamma(\lambda+n+1/2)}\,.
\end{align}
\item As we stated before, if $k\geq 1$, then the leading terms for $\rho \to \pm \pi/2$ are identical with those in case ({\it i}).  Hence
the only square-integrable functions (up to a normalization factor) are given again by $\Psi_\omega^{(1)}(\rho)$ with $\omega=\lambda+2\ell$, $\ell\in \mathbb{N}\cup \{0\}$, in Eq.~\eqref{transf1} and $\Psi_\omega^{(2)}(\rho)$ with $\omega=\lambda+2\ell+1$, $\ell\in \mathbb{N}\cup \{0\}$ in Eq.~\eqref{transf2}, where $\lambda = k+1/2$.
Equations~\eqref{transf1} and \eqref{transf2} are ill-defined for these values of $\omega$ as they stand, but by observing that, for $j\leq \ell$,
\begin{subequations}
\begin{align}
    \lim_{\omega\to \lambda+2\ell}H_1(\omega)h_1(j) & = \frac{(-1)^{\ell+1}\Gamma(\frac{1}{2})\Gamma(1+k+\ell)}{\Gamma(\ell+\frac{1}{2})}\,,\\
    \lim_{\omega\to \lambda+2\ell+1}H_2(\omega)h_2(j) & = \frac{(-1)^{\ell+1}\Gamma(\frac{3}{2})\Gamma(1+k+\ell)}{\Gamma(\ell+\frac{3}{2})}\,,
\end{align}
\end{subequations}
with these limits vanishing if $j \geq \ell+1$, we find that $\Psi_\omega^{(1)}(\rho)$ and $\Psi_\omega^{(2)}(\rho)$ behave like $(\cos\rho)^\lambda$ as $\rho\to \pm \pi/2$ for $\omega=\lambda+2\ell$ and $\omega=\lambda+2\ell+1$, respectively.
Thus, also in these cases, the Klein-Gordon normalized positive-frequency mode functions are given by
Eq.~\eqref{case1sol}.
For the case $\lambda=1/2$, the function $\Psi(\rho)$ in Eq.~\eqref{asympt} is square integrable for all $C_1$, $C_2$ and $\omega$. To treat this case we need to analyze the boundary conditions which give self-adjoint extensions of the operator $A$ given by Eq.~\eqref{OpA}.
This analysis will be given in Sec.~\ref{SAC}.
\item Here, we have $-1<2-2\lambda<1$, so the function $\Psi(\rho)$ in Eq.~\eqref{asymptotic} is square integrable for all values of $C_1$, $C_2$ and $\omega$.  Therefore, we need to determine the boundary conditions which give self-adjoint extensions of the operator $A$.  This task will be carried out in Sec.~\ref{SAC}.
\end{enumerate}

In this section we have seen that the leading behavior of the solutions to the eigenvalue equation for the adjoint $A^\dagger$ is uniquely determined if $\lambda \geq 3/2$. In these cases, the adjoint $A^\dagger$ turns out to be self-adjoint, as we discuss below.
On the other hand, for $1/2\leq\lambda<3/2$ all solutions are square integrable. In these cases we need to find suitable boundary conditions at $\rho=\pm \pi/2$ which give self-adjoint extensions of the operator $A$ in Eq.~\eqref{OpA}.  This task will be carried out next.

\subsection{Self-adjoint extensions of the operator $A$}\label{SAE}

Now we discuss the self-adjoint extensions of the operator $A$ defined in Eq.~\eqref{OpA} on the domain $\dom{A} = C_c^\infty(-\pi/2,\pi/2)$. The theory of self-adjoint extensions is due to Weyl~\cite{Weyl} and von~Neumann,~\cite{Neu} and a detailed summary of the general theory which we will briefly outline, can be found in the standard literature.\cite{reed,kom}

We start by finding the deficiency subspaces of the symmetric operator $A$, which are defined as the linear span of the (normalizable) solutions to the equations
\begin{align}\label{deficiencyEqs}
A^{\dagger}\Phi_{\pm}(\rho)=\pm 2i\Phi_{\pm}(\rho)\,.
\end{align} 
(The eigenvalues $\pm 2i$ may be replaced by any number $z \in \mathbb{C}\setminus \mathbb{R}$ and its complex conjugate $\overline{z}$.)
In other words, the positive and negative deficiency subspaces are given by
\begin{align}\label{defspaces}
\mathscr{K}_{\pm}:=&\,\mathrm{Ker}(A^{\dagger}\mp 2i\mathbb{I})\,,
\end{align}
where, $\mathbb{I}$ is the identity operator.
The dimensions of the subspaces $\mathscr{K}_\pm$, denoted by $n_\pm$, are called the deficiency indices.

An operator $\Omega$ is said to be closed if the \textit{graph} of $\Omega$, which is the set consisting of $(\Psi,\Omega\Psi)\in L^2[-\pi/2,\pi/2]\times L^2[-\pi/2,\pi/2]$ with $\Psi\in \dom{\Omega}$, is a closed set. The smallest closed extension of an operator is called its closure.  It is known that a densely-defined symmetric operator, such as the operator $A$, has a closure, $\bar{A} = (A^\dagger)^\dagger$ (see, e.g., Reed and Simon~\cite{reed-one}).  The domain of the adjoint $A^\dagger$ is~\cite{reed-one}
\begin{align}
    \dom{A^\dagger}=\dom{\bar{A}}\oplus \mathscr{K}_+ \oplus \mathscr{K}_{-}\,.
\end{align}
Suppose that $n_{\pm}$ are finite and $n_{+}=n_{-}$. Then,
a self-adjoint extension $A_U$ of $A$ is a restriction of $A^\dagger$ to a domain of the form~\cite{reed,kom}
\begin{align}
    \dom{A_U} = \dom{\bar{A}}\oplus \mathscr{S}\,,
\end{align}
where $\mathscr{S}$ is an $n_\pm$-dimensional subspace of $\mathscr{K}_+\oplus \mathscr{K}_{-}$ such that $\langle A^\dagger \Psi_1,\Psi_2\rangle = \langle \Psi_1,A^\dagger\Psi_2\rangle$ if $\Psi_1,\Psi_2 \in \dom{A_U}$. If $n_+\neq n_-$, then there are no self-adjoint extensions of $A$.

We first determine the deficiency indices by finding normalizable solutions of Eq.~\eqref{deficiencyEqs}. Notice that this equation is Eq.~\eqref{spatialKG} with $\omega^2 = \pm 2i$. 
Thus, each equation has two linearly independent solutions. The solutions in $\mathscr{K}_{+}$, if they are square integrable, are given by Eq.~\eqref{Solutions} with $\omega=1+ i$ as 
\begin{subequations}\label{defSolutions1}
\begin{align}
\Phi^{(1)}(\rho)&=c_1(\cos\rho)^{\lambda}\,F\left(\frac{\lambda+1+ i}{2},\frac{\lambda-1- i}{2};\frac{1}{2};\sin^{2}\rho\right)\,,\label{defSolutions1-1}\\
\Phi^{(2)}(\rho)&=c_2\sin\rho\,(\cos\rho)^{\lambda}\,F\left(\frac{\lambda+2+ i}{2},\frac{\lambda - i}{2};\frac{3}{2};\sin^{2}\rho\right)\,,\label{defSolutions1-2}
\end{align}
\end{subequations}
where the normalization constants $c_j\in \mathbb{R}$, $j=1,2$, are chosen so that
\begin{align}
    \langle \Phi^{(1)},\Phi^{(1)}\rangle = \langle\Phi^{(2)},\Phi^{(2)}\rangle = 1\,.
\end{align}
Note that $\langle \Phi^{(1)},\Phi^{(2)}\rangle = 0$ because $\Phi^{(1)}$ and $\Phi^{(2)}$ are even and odd functions, respectively.
We have $\Phi^{(1)},\Phi^{(2)}\in \mathscr{K}_{+}$ and $\overline{\Phi^{(1)}},\overline{\Phi^{(2)}}\in \mathscr{K}_{-}$ if $\Phi^{(1)}$ and $\Phi^{(2)}$ are square integrable.

From the analysis preceding the solutions given by Eq.~\eqref{case1sol}, we know that if $\lambda\geq 3/2$, then Eq.~\eqref{spatialKG} has square-integrable solutions only if $\omega=\omega^{\mathrm{I}}_n=\lambda+n$, with $n\in\mathbb{N}\cup\{0\}$. This means that there are no square-integrable solutions to Eq.~\eqref{deficiencyEqs} for $\omega=1\pm i$. Then it follows that the spaces $\mathscr{K}_\pm$ defined by Eq.~\eqref{defspaces} are both zero-dimensional, \textit{i.e.}\  $n_\pm=0$, and so there is only one self-adjoint extension for $A$, namely, its closure $\bar{A}(=A^\dagger)$.  In this case the spectrum of $\bar{A}$ is discrete. That is, the eigenfunctions $\Psi_n^{\mathrm{I}}$ form an orthonormal basis for $L^2[-\pi/2,\pi/2]$.  Note also that the eigenvalues $(\omega_n^{\mathrm{I}})^2$ are all positive. Thus, the quantum field theory and the vacuum state can be constructed using the mode functions $\phi_n^{\mathrm{I}}(t,\rho)$ following the general procedure outlined in Appendix~\ref{App-static-spacetime}.

For the cases $1/2\leq\lambda<3/2$ both solutions in Eq.~\eqref{defSolutions1} are square integrable and, hence, in $\dom{A^\dagger}$. This follows from the fact that these functions have the same asymptotic behavior as that of Eq.~\eqref{asymptotic} if $1/2\leq\lambda<3/2$, and of Eq.~\eqref{asympt} if $\lambda=1/2$, which are both square integrable for any value of $\omega$, in particular, for $\omega=1\pm i$. Hence, we have $n_{+}=n_{-}=2$.  Thus, the self-adjoint extensions of $A$ are characterized by the subspaces $\mathscr{S}$ of $\mathscr{K}_+\oplus \mathscr{K}_{-}$ on which the operator $A^\dagger$ is symmetric.

Let $\Phi^{(i)} = \Phi_{+}^{(i)} + \Phi_{-}^{(i)}$, $i=1,2$, where $\Phi_{+}^{(i)}\in \mathscr{K}_+$ and $\Phi_{-}^{(i)} \in \mathscr{K}_{-}$. Then,
the condition
$\langle \Phi^{(i)},A^\dagger\Phi^{(j)}\rangle =  \langle A^\dagger\Phi^{(i)}, \Phi^{(j)}\rangle$ implies $\langle \Phi_{+}^{(i)},\Phi_{+}^{(j)}\rangle = \langle \Phi_{-}^{(i)},\Phi_{-}^{(j)}\rangle$.
Thus, if $\Phi^{(i)}$, $i=1,2$ is in $\mathscr{S}$, then $\Phi_{-}^{(i)} = U\Phi_{+}^{(i)}$, where $U: \mathscr{K}_+ \to \mathscr{K}_{-}$ is a unitary map.  Hence, the self-adjoint extensions are characterized by the following $2\times 2$ unitary map $U$:
\begin{align}\label{Uaction}
U\Phi_{+}:=U\begin{pmatrix}
\Phi^{(1)}\\[.5em]
\Phi^{(2)}
\end{pmatrix}=\begin{pmatrix}
u_{11}\overline{\Phi^{(1)}}+u_{12}\overline{\Phi^{(2)}}\\[.5em]
u_{21}\overline{\Phi^{(1)}}+u_{22}\overline{\Phi^{(2)}}
\end{pmatrix}\,,
\end{align}
where the $2\times 2$ matrix,
\begin{equation}\label{Uaction-matrix}
    U_{\mathrm{M}} = \begin{pmatrix} u_{11} & u_{12} \\ u_{21} & u_{22}\end{pmatrix}\,,
\end{equation}
is unitary.
The domain of the self-adjoint extension $A_U$ is given by
\begin{align}\label{DomExtA}
\dom{A_U}:=\left\{\Phi+\Phi_{+}+U\Phi_{+}\,\left|\,\Phi\in\dom{\bar{A}},\,\Phi_{+}\in\mathscr{K}_{+}\right.\right\}\,,
\end{align}
where the operator $A_U$ acts on this domain as
\begin{align}\label{AU-acting}
A_U(\Phi+\Phi_{+}+U\Phi_{+}) =A^{\dagger}\Phi+2i\Phi_{+}-2i U\Phi_{+}\,.
\end{align}

Although Eqs.~\eqref{Uaction}, \eqref{DomExtA} and \eqref{AU-acting} give all self-adjoint extensions of the operator $A$, it will be more convenient to describe them in terms of boundary values of the functions in the domain of $A_U$.  Since the functions in the deficiency subspace $\mathscr{K}_+\oplus \mathscr{K}_{-}$ behave either like $(\cos\rho)^{1-\lambda}$ or $(\cos\rho)^\lambda$ for $1/2 < \lambda < 3/2$ and either like $(\cos\rho)^{1/2}$ or $(\cos\rho)^{1/2}\ln(\cos^2\rho)$ for $\lambda=1/2$ as $\rho\to \pm \pi/2$, we define the following quantities in order to extract the boundary behavior of these functions:
\begin{align}\label{weightedsol}
\widetilde{\Psi}^{(\lambda)}(\rho):=(\cos\rho)^{\lambda-1}\Psi(\rho)\,,\hspace{.5cm}D\widetilde{\Psi}^{(\lambda)}(\rho):=(\cos\rho)^{2-2\lambda}\dfrac{d\ }{d\rho}\left((\cos\rho)^{\lambda-1}\Psi(\rho)\right)\,,
\end{align}
for the case $1/2<\lambda<3/2$, and
\begin{align}\label{weightedsol2}
\begin{split} \widetilde{\Psi}^{(1/2)}(\rho) & :=\frac{(\cos\rho)^{-1/2}}{\ln(\cos^2\rho)-1}\Psi(\rho)\,,\\
D\widetilde{\Psi}^{(1/2)}(\rho) & :=(\cos\rho)[\ln(\cos^2\rho)-1]^2\dfrac{d\ }{d\rho}\left(\frac{(\cos\rho)^{-1/2}}{\ln(\cos^2\rho)-1}\Psi(\rho)\right)\,,
\end{split}
\end{align}
if $\lambda=1/2$. At least one of the boundary values $\widetilde{\Psi}^{(\lambda)}(\pm\pi/2)$ and $D\widetilde{\Psi}^{(\lambda)}(\pm\pi/2)$ is non-zero for $\Psi\in \mathscr{K}_+\oplus \mathscr{K}_{-}$ (except for the null function), and they vanish if $\Psi\in \dom{\bar{A}}$ as shown in Appendix~\ref{App-A-bar-boundary}.  Note also that for any given set of four values $\widetilde{\Psi}^{(\lambda)}(\pm\pi/2)$ and $D\widetilde{\Psi}^{(\lambda)}(\pm\pi/2)$, one can find a function $\Psi \in \dom{A^\dagger}$ which has these boundary values.
(For example, for $1/2 < \lambda < 3/2$ a smooth function $\Psi$ satisfying $\Psi(\rho) = (\cos\rho)^{1-\lambda}$ for $\pi/4 \leq \rho < \pi/2$ and $\Psi(\rho)=0$ for $-\pi/2 \leq \rho \leq 0$ is in $\dom{A^\dagger}$ and has $\widetilde{\Psi}^{(\lambda)}(\pi/2) = 1$ and $\widetilde{\Psi}^{(\lambda)}(-\pi/2) = D\widetilde{\Psi}^{(\lambda)}(\pm\pi/2) = 0$.)
These facts imply that the four-dimensional vector $(\widetilde{\Psi}^{(\lambda)}(-\pi/2),\widetilde{\Psi}^{(\lambda)}(\pi/2), D\widetilde{\Psi}^{(\lambda)}(-\pi/2),D\widetilde{\Psi}^{(\lambda)}(\pi/2))$ uniquely determines a function $\Psi \in \mathscr{K}_+\oplus\mathscr{K}_{-}$.  Thus, a self-adjoint extension of $A$ can be characterized by a two-dimensional subspace of the four-dimensional vector space consisting of these vectors which characterizes a subspace $\mathscr{S} \subset \mathscr{K}_+\oplus \mathscr{K}_-$, for which the operator $A^\dagger$ is symmetric.  We now find such subspaces of this four-dimensional vector space.

For either $\lambda=1/2$ or $1/2 < \lambda < 3/2$, if $\Psi_1,\Psi_2 \in \mathscr{K}_+ \oplus \mathscr{K}_-$, we have from Eq.~\eqref{boundary-for-A-dagger}
\begin{align}
    &\langle A^\dagger \Psi_1,\Psi_2\rangle - \langle\Psi_1,A^\dagger\Psi_2\rangle\nonumber \\ &=\overline{\widetilde{\Psi}_1^{(\lambda)}\left(\pi/2\right)}D\widetilde{\Psi}_2^{(\lambda)}\left(\pi/2\right) - \overline{D\widetilde{\Psi}_1^{(\lambda)}\left(\pi/2\right)}\widetilde{\Psi}_2^{(\lambda)}\left(\pi/2\right) \nonumber \\
    & \ \ \ - \left[\overline{\widetilde{\Psi}_1^{(\lambda)}\left(-\pi/2\right)}D\widetilde{\Psi}_2^{(\lambda)}\left(-\pi/2\right) - \overline{D\widetilde{\Psi}_1^{(\lambda)}\left(-\pi/2\right)}\widetilde{\Psi}_2^{(\lambda)}\left(-\pi/2\right)\right]\,.
    \end{align}
    Thus, the condition $\langle A^\dagger \Psi_1,\Psi_2\rangle - \langle\Psi_1,A^\dagger\Psi_2\rangle = 0$ can be written as
\begin{align}
    & \overline{\left[D\widetilde{\Psi}_1^{(\lambda)}\left(\pi/2\right) - i\widetilde{\Psi}_1^{(\lambda)}\left(\pi/2\right)\right]}\left[D\widetilde{\Psi}_2^{(\lambda)}\left(\pi/2\right)- i\widetilde{\Psi}_2^{(\lambda)}\left(\pi/2\right)\right]\notag \\
    & + \overline{\left[D\widetilde{\Psi}_1^{(\lambda)}\left(-\pi/2\right) + i\widetilde{\Psi}_1^{(\lambda)}\left(-\pi/2\right)\right]}\left[D\widetilde{\Psi}_2^{(\lambda)}\left(-\pi/2\right)+ i\widetilde{\Psi}_2^{(\lambda)}\left(-\pi/2\right)\right]\nonumber \\
    & = \overline{\left[D\widetilde{\Psi}_1^{(\lambda)}\left(\pi/2\right) + i\widetilde{\Psi}_1^{(\lambda)}\left(\pi/2\right)\right]}\left[D\widetilde{\Psi}_2^{(\lambda)}\left(\pi/2\right)+ i\widetilde{\Psi}_2^{(\lambda)}\left(\pi/2\right)\right] \notag \\
    & \quad + \overline{\left[D\widetilde{\Psi}_1^{(\lambda)}\left(-\pi/2\right) - i\widetilde{\Psi}_1^{(\lambda)}\left(-\pi/2\right)\right]}\left[D\widetilde{\Psi}_2^{(\lambda)}\left(-\pi/2\right)- i\widetilde{\Psi}_2^{(\lambda)}\left(-\pi/2\right)\right]\,.
\end{align}
This relation is equivalent to
\begin{align}\label{first-self-adjoint-BC}
    \begin{pmatrix} D\widetilde{\Psi}^{(\lambda)}\left(\pi/2\right) - i \widetilde{\Psi}^{(\lambda)}\left(\pi/2\right) \\ D\widetilde{\Psi}^{(\lambda)}\left(-\pi/2\right) + i \widetilde{\Psi}^{(\lambda)}\left(-\pi/2\right)\end{pmatrix} =  \mathcal{U}\begin{pmatrix} D\widetilde{\Psi}^{(\lambda)}\left(\pi/2\right) + i \widetilde{\Psi}^{(\lambda)}\left(\pi/2\right) \\ D\widetilde{\Psi}^{(\lambda)}\left(-\pi/2\right) - i \widetilde{\Psi}^{(\lambda)}\left(-\pi/2\right)\end{pmatrix}\,,
\end{align}
for all $\Psi\in \mathscr{S}$ for a fixed $2\times 2$ unitary matrix $\mathcal{U}$.  (This relation specifying the two-dimensional subspace $\mathscr{S}$ of $\mathscr{K}_+\oplus \mathscr{K}_{-}$  is analogous to that for a free quantum particle in a box.\cite{bonneau})
The unitary matrices $\mathcal{U}$ above and $U_{\mathrm{M}}$ in Eq.~\eqref{Uaction-matrix} both characterize a subspace $\mathscr{S}$. We find a map $U_{\mathrm{M}}\mapsto \mathcal{U}$ which identifies the unitary matrix $\mathcal{U}$ that specifies the same subspace $\mathscr{S}$ as the unitary matrix $U_{\mathrm{M}}$ in Appendix~\ref{appendix}.

Note that the condition~\eqref{first-self-adjoint-BC} can also be expressed as
\begin{align}\label{SABC}
(\mathbb{I}-\mathcal{U})\begin{pmatrix}
D\widetilde{\Psi}^{(\lambda)}\left(\pi/2\right)\\[.5em]
D\widetilde{\Psi}^{(\lambda)}\left(-\pi/2\right)
\end{pmatrix}
=i(\mathbb{I}+\mathcal{U})\begin{pmatrix}
\widetilde{\Psi}^{(\lambda)}\left(\pi/2\right)\\[.5em]
-\widetilde{\Psi}^{(\lambda)}\left(-\pi/2\right)
\end{pmatrix}\,.
\end{align}

Thus, we have a family of self-adjoint operators $A_U$ parametrized by the four entries of the unitary matrix $\mathcal{U}$. (We emphasize that the operator $A_U$ is the same differential operator for all $\mathcal{U}$.)  The quantities $\widetilde{\Psi}^{(\lambda)}(\pm\pi/2)$ and $D\widetilde{\Psi}^{(\lambda)}(\pm\pi/2)$ for the general eigenfunctions $C_1\Psi_\omega^{(1)}(\rho)+C_2\Psi_\omega^{(2)}(\rho)$, where $\Psi_\omega^{(1)}(\rho)$ and $\Psi_\omega^{(2)}(\rho)$ are given by Eq.~\eqref{expansion}, are linear in $C_1$ and $C_2$ with their coefficients being functions of the frequency $\omega$. Hence, Eq.~\eqref{SABC} gives two equations linear in $C_1$ and $C_2$. The condition for the existence of non-trivial solutions for $C_1$ and $C_2$ gives the spectrum of $\omega^2$ for each unitary matrix $\mathcal{U}$.  It is known that the spectrum of the self-adjoint extension of a second-order differential operator with deficiency indices $n_\pm=2$ is discrete.\cite{bonneau,naimarkbook}  Therefore, the eigenfunctions of the operator $A_U$ satisfying the boundary conditions~\eqref{SABC} form a basis for the Hilbert space $L^2[-\pi/2,\pi/2]$.  If all eigenvalues $\omega^2$ are positive, then one can follow the standard procedure to quantize this theory with a stationary vacuum state as outlined in Appendix~\ref{App-static-spacetime}. We provide an example of boundary conditions with negative eigenvalues for $\omega^2$ in Appendix~\ref{App-negative-eigenvalues}.

Next, let us write the boundary conditions~\eqref{SABC} in a more familiar form.
First let us consider the case for which the matrix $\mathbb{I}-\mathcal{U}$ is regular. In this case the matrix $i(\mathbb{I}-\mathcal{U})^{-1}(\mathbb{I}+\mathcal{U})$ is Hermitian. Hence, Eq.~\eqref{SABC} can be written as
\begin{subequations}\label{BC1}
\begin{align}
D\widetilde{\Psi}^{(\lambda)}\left(\pi/2\right)&=\alpha\,\widetilde{\Psi}^{(\lambda)}\left(\pi/2\right)-\beta\,\widetilde{\Psi}^{(\lambda)}\left(-\pi/2\right)\,,\label{BC1-1}\\
D\widetilde{\Psi}^{(\lambda)}\left(-\pi/2\right)&=\overline{\beta}\,\widetilde{\Psi}^{(\lambda)}\left(\pi/2\right)+\gamma\,\widetilde{\Psi}^{(\lambda)}\left(-\pi/2\right)\,,\label{BC1-2}
\end{align}
\end{subequations}
where $\alpha, \gamma\in\mathbb{R}$ and $\beta\in\mathbb{C}$. Notice that if $\mathcal{U}$ is a diagonal matrix, then $\beta=0$ and Eq.~\eqref{BC1} reduces to what can be called generalized Robin boundary conditions. A special case of Eq.~\eqref{BC1} is given by $\alpha=\beta=\gamma=0$ (corresponding to $\mathcal{U} = -\mathbb{I}$) as
\begin{align}\label{NeumannBC}
D\widetilde{\Psi}^{(\lambda)}\left(\pi/2\right)=D\widetilde{\Psi}^{(\lambda)}\left(-\pi/2\right)=0\,,
\end{align}
which we call the generalized Neumann boundary condition.  If $\mathbb{I}+\mathcal{U}$ is invertible, then the matrix $i(\mathbb{I}+\mathcal{U})^{-1}(\mathbb{I}-\mathcal{U})$ is Hermitian, and Eq.~\eqref{SABC} can be given as
 \begin{subequations}\label{InvBC1}
\begin{align}
\widetilde{\Psi}^{(\lambda)}\left(\pi/2\right)&=a\,D\widetilde{\Psi}^{(\lambda)}\left(\pi/2\right)  -b\,D\widetilde{\Psi}^{(\lambda)}\left(-\pi/2\right)\,,\\
\widetilde{\Psi}^{(\lambda)}\left(-\pi/2\right)&=\overline{b}\,D\widetilde{\Psi}^{(\lambda)}\left(\pi/2\right)+c\,D\widetilde{\Psi}^{(\lambda)}\left(-\pi/2\right)\,,
\end{align}
\end{subequations}
where $a,c\in\mathbb{R}$ and $b\in\mathbb{C}$.
The special case with $a=b=c=0$ (corresponding to $\mathcal{U}=\mathbb{I}$) is
\begin{align}\label{DirichletBC}
\widetilde{\Psi}^{(\lambda)}\left(\pi/2\right)=\widetilde{\Psi}^{(\lambda)}\left(-\pi/2\right)=0\,,
\end{align}
which we call the generalized Dirichlet boundary condition. 

If the matrices $\mathbb{I}\pm \mathcal{U}$ are both singular, \textit{i.e.}\ if the eigenvalues of $\mathcal{U}$ are $\pm 1$, then it is a Pauli spin matrix and can be parametrized as
\begin{align}
    \mathcal{U} = \begin{pmatrix} \cos 2\theta & e^{i\varphi}\sin 2\theta \\ e^{-i\varphi}\sin 2\theta & - \cos 2\theta\end{pmatrix}\,,
\end{align}
where $\theta,\varphi\in\mathbb{R}$.  Then Eq.~\eqref{SABC} reduces to
\begin{subequations}\label{GenAutBC}
\begin{align}
\widetilde{\Psi}^{(\lambda)}\left(\pi/2\right)\cos\theta & = \widetilde{\Psi}^{(\lambda)}\left(-\pi/2\right)e^{i\varphi}\sin\theta\,,\label{GenAutBC-1} \\
    D\widetilde{\Psi}^{(\lambda)}\left(\pi/2\right)\sin\theta & = D\widetilde{\Psi}^{(\lambda)}\left(-\pi/2\right)e^{i\varphi}\cos\theta\,.\label{GenAutBC-2}
\end{align}
\end{subequations}
The special cases with $\theta= \pi/4$ gives what can be called the generalized automorphic boundary conditions.  If we set $\varphi=0$ as well, then we have the periodic boundary condition.  The special case with $\theta=\pi/2$ reads
\begin{align}\label{K0}
\widetilde{\Psi}^{(\lambda)}\left(\pi/2\right) & =D\widetilde{\Psi}^{(\lambda)}\left(-\pi/2\right)=0\,,
\end{align}
whereas the case with $\theta=0$ reads
\begin{align}\label{Kinft}
\widetilde{\Psi}^{(\lambda)}\left(-\pi/2\right) & =D\widetilde{\Psi}^{(\lambda)}\left(\pi/2\right)=0\,.
\end{align}
These boundary conditions will be called the mixed boundary conditions.

Most of the boundary conditions given by Eq.~\eqref{SABC} will result in a rather complicated spectrum of the frequency $\omega$. This is related to the fact that they are not invariant under the symmetry group $\widetilde{\mathrm{SL}}(2,\mathbb{R})$ of $\mathrm{AdS}_2$. Next we identify all boundary conditions among those given by Eq.~\eqref{SABC} that are invariant under this symmetry group.

\subsection{The invariant self-adjoint boundary conditions}\label{InvBC}

In order to determine which of the (positive) self-adjoint extensions of the operator $A$ result in a representation of $\widetilde{\mathrm{SL}}(2,\mathbb{R})$, we first find the boundary conditions which are invariant under the infinitesimal action of the group. 
The action of the operators $L_{\pm}$ on the mode functions of the form $\phi(t,\rho)=\Psi_\omega(\rho)e^{-i\omega t}$ in our local coordinates can be found by using the definitions in Eqs.~\eqref{KillingVF} and \eqref{ladder} as
\begin{align}\label{ladder2}
\mp i L_{\pm}\phi(t,\rho)= e^{-i(\omega\pm 1)t}\left(\cos\rho\frac{d\ }{d\rho}\Psi_\omega(\rho)\mp\omega\sin\rho\Psi_\omega(\rho)\right)\,.
\end{align}
Thus, at $t=0$ the function $\Psi_\omega(\rho)$ and its derivative transform under the action of $\mp i L_\pm$ as follows:
\begin{subequations}\label{transf-Psi}
\begin{align}
    \delta_\pm \Psi_\omega(\rho) & = \cos\rho \frac{d\ }{d\rho}\Psi_\omega(\rho)\mp \omega \sin\rho\,\Psi_\omega(\rho)\,,\label{transf-Psi-1}\\
    \delta_\pm \left( \frac{d\ }{d\rho}\Psi_\omega(\rho)\right) & = (-1\mp\omega)\sin\rho\frac{d\ }{d\rho}\Psi_\omega(\rho) + (-\omega^2\mp \omega)\cos\rho\Psi_\omega(\rho)\nonumber \\
    &\ \ \ - \frac{\lambda(1-\lambda)}{\cos\rho}\Psi_\omega(\rho)\,, \label{transf-Psi-2}
\end{align}
\end{subequations}
where we have used the Klein-Gordon equation~\eqref{spatialKG} to find Eq.~\eqref{transf-Psi-2}.

First we examine the cases with $1/2 < \lambda < 3/2$.  Using the definitions of $\widetilde{\Psi}_\omega^{(\lambda)}$ and $D \widetilde{\Psi}_\omega^{(\lambda)}$ in Eq.~\eqref{weightedsol}, we find from Eq.~\eqref{transf-Psi}
\begin{subequations}
\begin{align}
    \delta_{-}\widetilde{\Psi}_\omega^{(\lambda)}(\rho) & = (\omega-1+\lambda)\sin\rho\,\widetilde{\Psi}^{(\lambda)}(\rho)+(\cos\rho)^{2\lambda-1}D\widetilde{\Psi}_\omega^{(\lambda)}(\rho)\,,\\
    \delta_{-} D\widetilde{\Psi}_\omega^{(\lambda)}(\rho) & = (\omega-\lambda)\sin\rho\,D\widetilde{\Psi}_\omega^{(\lambda)}(\rho) +\left[\lambda(\lambda-1)-\omega(\omega-1)\right](\cos\rho)^{3-2\lambda}\widetilde{\Psi}_\omega^{(\lambda)}(\rho)\,.
\end{align}
\end{subequations}
The formulas for $\delta_+$ are obtained from these by letting $\omega \mapsto -\omega$. Then
\begin{subequations}
\begin{align}
    \delta_-\widetilde{\Psi}_\omega^{(\lambda)}\left(\pm\pi/2\right) & = \pm (\omega-1+\lambda)\widetilde{\Psi}_\omega^{(\lambda)}\left(\pm\pi/2\right)\,,\\
    \delta_-D\widetilde{\Psi}_\omega^{(\lambda)}\left(\pm\pi/2\right) & = \pm (\omega-\lambda) D\widetilde{\Psi}_\omega^{(\lambda)}\left(\pm\pi/2\right)\,.
\end{align}
\end{subequations}

Now, if $\Psi_{\omega_1}$ and $\Psi_{\omega_2}$ are eigenfunctions with the same boundary condition with $\omega_1,\omega_2\in \mathbb{R}$, then
\begin{align}\label{symmetry-condition}
    \langle &A_U\Psi_{\omega_1},\Psi_{\omega_2}\rangle - \langle \Psi_{\omega_1},A_U\Psi_{\omega_2}\rangle \nonumber \\  &=\left[\overline{\widetilde{\Psi}_{\omega_1}\left(\pi/2\right)}D\widetilde{\Psi}_{\omega_2}\left(\pi/2\right) - \overline{D\widetilde{\Psi}_{\omega_1}\left(\pi/2\right)}\widetilde{\Psi}_{\omega_2}\left(\pi/2\right)\right] \nonumber \\
    & \ \ \ - \left[\overline{\widetilde{\Psi}_{\omega_1}\left(-\pi/2\right)}D\widetilde{\Psi}_{\omega_2}\left(-\pi/2\right) - \overline{D\widetilde{\Psi}_{\omega_1}\left(-\pi/2\right)}\widetilde{\Psi}_{\omega_2}\left(-\pi/2\right)\right]\nonumber \\
    & = 0\,.
\end{align}
If this boundary condition is invariant under the transformation $\delta^{(\lambda)}_{-}$, then $\delta_{-}\Psi_\omega$ must satisfy the same boundary condition as $\Psi_\omega$.  This implies
\begin{equation}
    \langle A_U\delta_{-}\Psi_{\omega_1},\Psi_{\omega_2}\rangle - \langle \delta_{-}\Psi_{\omega_1},A_U\Psi_{\omega_2}\rangle = 0\,.
\end{equation}
Then,
\begin{align}\label{transform-of-symmetry-condition}
&  \left[\langle A_U\delta_{-}\Psi_{\omega_1},\Psi_{\omega_2}\rangle - \langle \delta_{-}\Psi_{\omega_1},A_U\Psi_{\omega_2}\rangle\right] + 
\left[\langle A_U\Psi_{\omega_1},\delta_{-}\Psi_{\omega_2}\rangle - \langle \Psi_{\omega_1},A_U\delta_{-}\Psi_{\omega_2}\rangle\right]\notag \\
& = (\omega_1+\omega_2-1)\left[\overline{\widetilde{\Psi}^{(\lambda)}_{\omega_1}\left(\pi/2\right)}D\widetilde{\Psi}^{(\lambda)}_{\omega_2}\left(\pi/2\right) - \overline{D\widetilde{\Psi}^{(\lambda)}_{\omega_1}\left(\pi/2\right)}\widetilde{\Psi}^{(\lambda)}_{\omega_2}\left(\pi/2\right)\right] \nonumber \\
    & \ \ \ +(\omega_1+\omega_2-1) \left[\overline{\widetilde{\Psi}^{(\lambda)}_{\omega_1}\left(-\pi/2\right)}D\widetilde{\Psi}^{(\lambda)}_{\omega_2}\left(-\pi/2\right) - \overline{D\widetilde{\Psi}^{(\lambda)}_{\omega_1}\left(-\pi/2\right)}\widetilde{\Psi}^{(\lambda)}_{\omega_2}\left(-\pi/2\right)\right]\nonumber \\
    & = 0\,.
\end{align}
Equations~\eqref{symmetry-condition} and \eqref{transform-of-symmetry-condition} are compatible with each other if and only if
\begin{align}
& \overline{\widetilde{\Psi}^{(\lambda)}_{\omega_1}\left(\pi/2\right)}D\widetilde{\Psi}^{(\lambda)}_{\omega_2}\left(\pi/2\right) - \overline{D\widetilde{\Psi}^{(\lambda)}_{\omega_1}\left(\pi/2\right)}\widetilde{\Psi}^{(\lambda)}_{\omega_2}\left(\pi/2\right) \notag \\
& = \overline{\widetilde{\Psi}^{(\lambda)}_{\omega_1}\left(-\pi/2\right)}D\widetilde{\Psi}^{(\lambda)}_{\omega_2}\left(-\pi/2\right) - \overline{D\widetilde{\Psi}^{(\lambda)}_{\omega_1}\left(-\pi/2\right)}\widetilde{\Psi}^{(\lambda)}_{\omega_2}\left(-\pi/2\right)\notag \\
&      = 0\,,
\end{align}
for all pairs $\{\Psi_{\omega_1},\Psi_{\omega_2}\}$ such that $\omega_1+\omega_2 \neq 1$ (and there are infinitely many such pairs).
This implies that the unitary matrix $\mathcal{U}$ must be diagonal for the boundary condition to be invariant under the $\widetilde{\mathrm{SL}}(2,\mathbb{R})$ transformations.  That is,
\begin{align}\label{simpler-BC}
    (1-e^{i\alpha_\pm})D\widetilde{\Psi}_\omega^{(\lambda)}\left(\pm\pi/2\right)  = \pm i(1+e^{i\alpha_{\pm}})\widetilde{\Psi}_\omega^{(\lambda)}\left(\pm\pi/2\right)\,,
\end{align}
where $\alpha_{\pm}\in \mathbb{R}$.  The $L_{-}$-transformation of these equations gives 
\begin{align}\label{transform-simpler-BC}
    \pm(\omega-\lambda)(1-e^{i\alpha_\pm})D\widetilde{\Psi}_\omega^{(\lambda)}\left(\pm\pi/2\right)  = i(\omega-1+\lambda) (1+e^{i\alpha_{\pm}})\widetilde{\Psi}_\omega^{(\lambda)}\left(\pm\pi/2\right)\,.
\end{align}
Equations~\eqref{simpler-BC} and \eqref{transform-simpler-BC} are compatible with each other if and only if $e^{i\alpha_{\pm}} = \pm 1$, \textit{i.e.}\ if and only if $\widetilde{\Psi}^{(\lambda)}(\pi/2)=0$ or $D\widetilde{\Psi}^{(\lambda)}(\pi/2)=0$, and $\widetilde{\Psi}^{(\lambda)}(-\pi/2)=0$ or $D\widetilde{\Psi}^{(\lambda)}(-\pi/2)=0$.  Thus, the only $\widetilde{\mathrm{SL}}(2,\mathbb{R})$-invariant boundary conditions are the generalized Dirichlet and Neumann boundary conditions, given by Eqs.~\eqref{DirichletBC} and \eqref{NeumannBC}, respectively, and the mixed boundary conditions given by Eqs,~\eqref{K0} and \eqref{Kinft}.  It can readily be seen that these boundary conditions are also invariant under the $L_{+}$-transformation, and hence under all $\widetilde{\mathrm{SL}}(2,\mathbb{R})$ transformations.

Next we turn to the case $\lambda =1/2$.  The transformation of $\Psi_\omega$ and its derivative under the infinitesimal group action is given by Eq.~\eqref{transf-Psi} with $\lambda=1/2$.  From the definitions of $\widetilde{\Psi}^{(1/2)}$ and $D\widetilde{\Psi}^{(1/2)}$ given by Eq.~\eqref{weightedsol2}, we find
\begin{subequations}
\begin{align}
    \delta_{-}\widetilde{\Psi}_\omega^{(1/2)}(\rho) & = \left(\omega-\frac{1}{2}- \frac{2}{\ln(\cos^2\rho)-1}\right)\sin\rho\,\widetilde{\Psi}_\omega^{(1/2)}(\rho) \nonumber \\
    &\;\;\;\; +\frac{1}{\left[\ln(\cos^2\rho)-1\right]^2}D\widetilde{\Psi}_\omega^{(1/2)}(\rho)\,,\\
    \delta_{-}D\widetilde{\Psi}_\omega^{(1/2)}(\rho) & = \left\{ -4 + \cos^2\rho\left[\left(-\omega^2+\omega-\frac{1}{4}\right)\left[\ln(\cos^2\rho)-1\right]^2 +4\right]\right\}\widetilde{\Psi}_\omega^{(1/2)}(\rho)\nonumber \\
    & \;\;\;\; +\left( \omega-\frac{1}{2} + \frac{2}{\ln(\cos^2\rho)-1}\right)\sin\rho\,D\widetilde{\Psi}_\omega^{(1/2)}(\rho)\,.
\end{align}
\end{subequations}
Thus, we obtain
\begin{subequations}
\begin{align}
    \delta_{-}\widetilde{\Psi}_\omega^{(1/2)}\left(\pm\pi/2\right) & = \pm \left( \omega-1/2\right)\widetilde{\Psi}_\omega^{(1/2)}\left(\pm\pi/2\right)\,,\\
    \delta_{-}D\widetilde{\Psi}_\omega^{(1/2)}\left(\pm\pi/2\right) & = \pm\left( \omega-1/2\right)D\widetilde{\Psi}_\omega^{(1/2)}\left(\pm\pi/2\right) - 4\widetilde{\Psi}_\omega^{(1/2)}\left(\pm\pi/2\right)\,.
\end{align}
\end{subequations}
It can be shown as in the cases with $1/2 < \lambda < 3/2$ that, for the boundary condition to be invariant under the $L_{-}$-transformation, the matrix $\mathcal{U}$ must be diagonal.  Thus, we have boundary conditions of the form given by Eq.~\eqref{simpler-BC} in this case as well.  Then the $L_{-}$-transformation gives
\begin{align}\label{transform-simpler-BC2}
     & \pm(\omega-1/2)(1-e^{i\alpha_\pm})D\widetilde{\Psi}_\omega^{(1/2)}\left(\pm\pi/2\right)\notag \\
     & = \left[i(\omega-1/2) (1+e^{i\alpha_\pm})\widetilde{\Psi}_\omega^{(\lambda)}\left(\pm\pi/2\right) + 4(1-e^{i\alpha_\pm})\right]\widetilde{\Psi}_\omega^{(1/2)}\left(\pm\pi/2\right)\,.
\end{align}
This equation is compatible with Eq.~\eqref{simpler-BC} if and only if $e^{i\alpha_{\pm}}=1$.  Thus, the only $\widetilde{\mathrm{SL}}(2,\mathbb{R})$-invariant boundary condition is the generalized Dirichlet boundary condition~\eqref{DirichletBC}.  This was expected because all invariant boundary conditions for $1/2<\lambda < 3/2$, the generalized Dirichlet, Neumann and mixed boundary conditions, tend to the generalized Dirichlet boundary condition in the limit $\lambda \to 1/2$.

\subsection{Invariant self-adjoint boundary conditions}\label{SAC}

Now that we have found the boundary conditions which are invariant under the group action $\widetilde{\mathrm{SL}}(2,\mathbb{R})$, we shall find the frequency spectrum and the corresponding mode functions for these boundary conditions.  Since all invariant boundary conditions are either the generalized Dirichlet or Neumann boundary condition at each boundary, it is convenient to use the solutions to Eq.~\eqref{spatialKG} satisfying one of these conditions at $\rho=\pi/2$.

Let us start with the cases with $1/2 < \lambda < 3/2$.  The solutions which satisfy the generalized Dirichlet or Neumann boundary condition at $\rho= \pi/2$ are
\begin{subequations}
\begin{align}
    \Psi_\omega^{(\mathrm{D},\lambda)}(\rho) & = (\cos\rho)^\lambda F\left( \lambda+\omega,\lambda-\omega;\lambda+\frac{1}{2}; \frac{1-\sin\rho}{2}\right)\,,\\
    \Psi_\omega^{(\mathrm{N},\lambda)}(\rho) & = (\cos\rho)^{1-\lambda}F\left( 1-\lambda+\omega,1-\lambda-\omega;\frac{3}{2}-\lambda; \frac{1-\sin\rho}{2}\right)\,,
\end{align}
\end{subequations}
respectively.  Note that the function $\Psi_\omega^{(\mathrm{N},\lambda)}(\rho)$ is obtained from $\Psi_\omega^{(\mathrm{D},\lambda)}(\rho)$ by letting $\lambda \mapsto 1-\lambda$.  The behavior of these functions for $\rho\to -\pi/2$ is given by 
\begin{subequations}\label{DN-at-right}
\begin{align}
    \Psi_\omega^{(\mathrm{D},\lambda)}(\rho) & \approx \frac{\cos\pi\omega}{\cos\pi\lambda}(\cos\rho)^\lambda + \frac{\Gamma(\lambda+\frac{1}{2})\Gamma(\lambda-\frac{1}{2})}{2^{1-2\lambda}\Gamma(\lambda+\omega)\Gamma(\lambda-\omega)}(\cos\rho)^{1-\lambda}\,,\label{DN-at-right-D}\\
    \Psi_\omega^{(\mathrm{N},\lambda)}(\rho) & \approx -\frac{\cos\pi \omega}{\cos\pi\lambda}(\cos\rho)^{1-\lambda}+ \frac{\Gamma(\frac{3}{2}-\lambda)\Gamma(\frac{1}{2}-\lambda)}{2^{2\lambda-1}\Gamma(1-\lambda+\omega)\Gamma(1-\lambda-\omega)}(\cos\rho)^\lambda\,.\label{DN-at-right-N}
    \end{align}
\end{subequations}

\noindent
\textbf{Dirichlet boundary condition.}
The function $\Psi_\omega^{(\mathrm{D},\lambda)}(\rho)$ given by Eq.~\eqref{DN-at-right-D} satisfies the generalized Dirichlet boundary condition also at $\rho=-\pi/2$ (with $\omega >0$) if and only if $\Gamma(\lambda-\omega)=\infty$, \textit{i.e.}\ if and only if $\omega=\omega_n^{\mathrm{I}}:= \lambda+n$, $n\in \mathbb{N}\cup\{0\}$.  The mode functions are given by Eq.~\eqref{case1sol}.

\smallskip

\noindent
\textbf{Neumann boundary condition.} We first discuss the cases with $\lambda\neq 1$.  The function $\Psi_\omega^{(\mathrm{N},\lambda)}(\rho)$ given by Eq.~\eqref{DN-at-right-N} satisfies the generalized Neumann boundary condition also at $\rho=-\pi/2$ (with $\omega > 0$) if and only if $\Gamma(1-\lambda-\omega)=\infty$ or $\Gamma(1-\lambda+\omega)=\infty$, \textit{i.e.}\ if and only if $\omega = 1-\lambda + n$, $n\in \mathbb{N}$, or $\omega=|1-\lambda|$.  For either case the positive-frequency mode functions are given by
\begin{align}\label{new-Neumann-modes}
 \phi^{\mathrm{II}}_n(t,\rho) & = N_n^{\mathrm{II}}(\cos\rho)^{1-\lambda}P_n^{(1/2-\lambda,1/2-\lambda)}(\sin\rho)e^{-i\omega_n^{\mathrm{II}} t}\,,
\end{align}
where $\omega_n^{\mathrm{II}} = 1-\lambda + n$, $n\in\mathbb{N}$, and $\omega_0^{\mathrm{II}} = |1-\lambda|$. The normalization constant such that $\langle \phi^{\mathrm{II}}_m,\phi^{\mathrm{II}}_n\rangle_{\mathrm{KG}} = \delta_{mn}$ can be found by using Eq.~\eqref{Jacobi-normalisation} as
\begin{align}
    N_n^{\mathrm{II}} = \frac{\sqrt{n!|\Gamma(2-2\lambda+n)|}}{2^{1-\lambda}\Gamma(3/2-\lambda+n)}\,.
\end{align}
Notice that $\omega_1-\omega_0 = 3-2\lambda\neq 1$ if $1 < \lambda < 3/2$.

For $\lambda=1$  we can write the positive-frequency solutions with the Neumann boundary condition as
\begin{align}
    \phi_n^{(\mathrm{II},\lambda=1)}(t,\rho) & = \begin{cases}\frac{1}{\sqrt{\pi n}}\sin n\rho\,e^{-int}\,, & n+1\in 2\mathbb{N}\,,\\
    \frac{1}{\sqrt{\pi n}}\cos n\rho\,e^{-int}\,, & n\in 2\mathbb{N}\,.\end{cases}
\end{align}
 We have $\langle \phi_m^{(\mathrm{II},\lambda=1)},\phi_n^{(\mathrm{II},\lambda=1)}\rangle_{\mathrm{KG}} = \delta_{mn}$ for $m,n\in\mathbb{N}$.
Note that there are solutions with $\omega=0$:
\begin{align}
 \phi_0^{(\mathrm{II},\lambda=1)}(t,\rho) & = A t + B\,,
 \end{align}
where $A$ and $B$ are constants.

\smallskip

\noindent
\textbf{Mixed boundary conditions.}  The function $\Psi_\omega^{(\mathrm{D},\lambda)}(\rho)$ in Eq.~\eqref{DN-at-right-D}, which satisfies the generalized Dirichlet boundary condition at $\rho=\pi/2$, will satisfy the generalized Neumann boundary condition at $\rho=-\pi/2$ if and only if $\cos\pi\omega=0$, \textit{i.e.}\  $\omega=n+1/2$, $n\in \mathbb{N}\cup \{0\}$ (for $\omega > 0$). Then
\begin{align}
    \Psi_\omega^{(\mathrm{D},\lambda)}(\rho)|_{\omega=n+1/2} & = (\cos\rho)^\lambda F\left(\lambda+n+\frac{1}{2}, \lambda-n-\frac{1}{2};\lambda+\frac{1}{2};\frac{1-\sin\rho}{2}\right)\nonumber \\
    & = (\cos\rho)^\lambda\left(\frac{1+\sin\rho}{2}\right)^{1/2-\lambda}F\left(1+n,-n;\lambda+\frac{1}{2};\frac{1-\sin\rho}{2}\right)\,.
\end{align}
The positive-frequency mode functions corresponding to this function are
\begin{align}
    \phi^{\mathrm{III}}_n(t,\rho) = N^{\mathrm{III}}_n \left(\cos\rho\right)^{\lambda}\left(1+\sin\rho\right)^{1/2-\lambda}P_n^{(\lambda-1/2,-\lambda+1/2)}(\sin\rho)e^{-i\omega_n^{\mathrm{III}}t}\,,
\end{align}
where $\omega_n^{\mathrm{III}}=n+1/2$ and where the normalization constant such that $\langle \phi_m^{\mathrm{III}},\phi_n^{\mathrm{III}}\rangle_{\mathrm{KG}} = \delta_{mn}$, found using Eq.~\eqref{Jacobi-normalisation}, is
\begin{align}
    N_n^{\mathrm{III}} = \frac{n!}{\sqrt{2\Gamma(\lambda+n+1/2)\Gamma(3/2-\lambda+n)}}\,.
\end{align}
The positive-frequency mode functions satisfying the generalized Dirichlet and Neumann boundary conditions at $\rho=-\pi/2$ and $\rho=\pi/2$, respectively, are
\begin{align}
     \phi^{\mathrm{IV}}_n(t,\rho) = N^{\mathrm{III}}_n \left(\cos\rho\right)^{\lambda}\left(1-\sin\rho\right)^{1/2-\lambda}P_n^{(-\lambda+1/2,\lambda-1/2)}(\sin\rho)e^{-i\omega_n^{\mathrm{III}}t}\,.
\end{align}
Since $P_n^{(\alpha,\beta)}(-x)=P_n^{(\beta,\alpha)}(x)$, we have
\begin{align}
    \phi_n^{\mathrm{IV}}(t,\rho) = \phi_n^{\mathrm{III}}(t,-\rho)\,,
\end{align}
as expected.

Next let us discuss the case $\lambda=1/2$. As we saw, the only invariant boundary condition in this case is the generalized Dirichlet boundary condition.  We have
\begin{align}
    \Psi_\omega^{(\mathrm{D},1/2)}(\rho) & = (\cos\rho)^{1/2} F\left( \frac{1}{2}+\omega,\frac{1}{2}-\omega;1;\frac{1-\sin\rho}{2}\right)\nonumber \\
    & = (\cos\rho)^{1/2} \mathrm{P}_{\omega-1/2}(\sin\rho)\,,
\end{align}
where $\mathrm{P}_\nu(x)$ is the Legendre function of the first kind. Then, since~\cite{abr}
\begin{align}
    \mathrm{P}_{\omega-1/2}(-x)= \frac{2}{\pi}\cos\pi \omega \,\mathrm{Q}_{\omega-1/2}(x) + \sin\pi\omega\mathrm{P}_{\omega-1/2}(x)\,,
\end{align}
where $\mathrm{Q}_\nu(x)$ is the Legendre function of the second kind with $\mathrm{Q}_\nu(x)\approx -\ln(1-x)/2$ as $x\to 1$, we must have $\cos\pi\omega = 0$ to have the generalized Dirichlet boundary condition at $\rho=-\pi/2$ as well.  Thus, we obtain the positive-frequency mode functions in this case as
\begin{align}
    \phi^{\mathrm{V}}_n(t,\rho) = \frac{1}{\sqrt{2}}(\cos\rho)^{1/2}\mathrm{P}_n(\sin\rho)e^{-i\omega_n^{\mathrm{V}} t}\,,
\end{align}
where $\omega_n^{\mathrm{V}} = n+1/2$.  Note that $\mathrm{P}_n(x) = P_n^{(0,0)}(x)$.  These mode functions are normalized so that $\langle\phi_m^{\mathrm{V}},\phi_n^{\mathrm{V}}\rangle = \delta_{mn}$.

\subsection{Boundary conditions from vanishing energy flux at the boundaries}\label{EnFlux}

The $\widetilde{\mathrm{SL}}(2,\mathbb{R})$-invariant boundary conditions and the solutions to Eq.~\eqref{spatialKG} found in Sec.~\ref{SAC} are identical with the results of Sakai and Tanii~\cite{sak} based on the requirement that the energy flux should vanish at the boundaries $\rho=\pm \pi/2$. (This requirement is analogous to that used in the higher-dimensional case.\cite{avis,breit,mezin}.) We briefly review this derivation. 

The stress-energy tensor with the conformal coupling constant $\beta$ is given by~\cite{sak}
\begin{align}
T_{\mu\nu}=\partial_{\mu}\phi\,\partial_{\nu}\phi-\frac{1}{2}g_{\mu\nu}g^{\kappa\sigma}\partial_{\kappa}\phi\,\partial_{\sigma}\phi-\frac{1}{2}g_{\mu\nu}M^{2}\phi^{2}+\frac{1}{2}\beta(g_{\mu\nu}\Box-\nabla_{\mu}\nabla_{\nu}+R_{\mu\nu})\phi^{2}\,,
\end{align}
where $R_{\mu\nu}= g_{\mu\nu}$ is the Ricci curvature, and $\nabla_{\mu}$ is the covariant derivative. The requirement of vanishing energy flux reads
\begin{align}\label{encond}
\left.\sqrt{-\mathrm{det}\,g}\,g^{1\mu}T_{\mu\nu}\Lambda_{0}^{\nu}\right|_{\rho=\pm\pi/2}=0\,,
\end{align}
where $\Lambda_{0}^{\mu}=\delta_0^\mu$ are the components of the Killing vector field $\Lambda_{0}$ in Eq.~\eqref{KillingVF} in global coordinates given by Eq.~\eqref{coords}.
From Eq.~\eqref{metric}, we have $g=\sec^{2}\rho\,\mathrm{diag}(-1,1)$, and therefore the only non-vanishing connection components are $\Gamma_{01}^{0}=\Gamma_{00}^{1}=\Gamma_{11}^{1}=\tan\rho$. By substituting the mode functions  given as in Eq.~\eqref{modeform} into Eq.~\eqref{encond}, we find
\begin{align}\label{sakai-tanii-condition}
\left.\left((1-2\beta)\frac{d\Psi_\omega(\rho)}{d\rho}+\beta\tan\rho\,\Psi_\omega(\rho)\right)\Psi_\omega(\rho)\right|_{\rho=\pm\pi/2}=0\,.
\end{align}

For $\lambda \geq 3/2$ we have $\Psi_\omega(\rho)\sim (\cos\rho)^\lambda$, and hence this condition is satisfied. For $1/2 < \lambda < 3/2$, the condition~\eqref{sakai-tanii-condition} becomes 
\begin{align}\label{sakai-tanii-2}
    &\left\{(1-2\beta)D \widetilde{\Psi}_\omega^{(\lambda)}(\rho)\right.\nonumber \\
    &+\left. \left[(3-2\lambda)\beta - (1-\lambda)\right](\cos\rho)^{1-2\lambda}\sin\rho\,\widetilde{\Psi}_\omega^{(\lambda)}(\rho)\right\}\widetilde{\Psi}_\omega^{(\lambda)}(\rho)\Big{|}_{\rho=\pm \pi/2} = 0\,,
\end{align}
where $\widetilde{\Psi}_\omega^{(\lambda)}(\rho)$ and $D\widetilde{\Psi}_\omega^{(\lambda)}(\rho)$ are defined by Eq.~\eqref{weightedsol}.  This condition is satisfied by the Dirichlet boundary condition $\widetilde{\Psi}^{(\lambda)}(\pm\pi/2)=0$ for all $\beta$.  If we choose
\begin{align}
    \beta = \frac{1-\lambda}{3-2\lambda}\,,
\end{align}
then the Neumann boundary condition $D\widetilde{\Psi}^{(\lambda)}(\pm\pi/2)=0$ and the mixed boundary conditions $D\widetilde{\Psi}^{(\lambda)}(\pi/2) = \widetilde{\Psi}^{(\lambda)}(-\pi/2)=0$ or $D\widetilde{\Psi}^{(\lambda)}(-\pi/2) = \widetilde{\Psi}^{(\lambda)}(\pi/2)=0$ satisfy the condition~\eqref{sakai-tanii-2}.

For $\lambda=1/2$, Eq.~\eqref{sakai-tanii-condition} reads
\begin{align}
    & \Bigg{\{}(1-2\beta)D\widetilde{\Psi}^{(\lambda)}(\rho)  + \Bigg{[}\left(2\beta-\frac{1}{2}\right)\left[\ln(\cos^2\rho)-1\right]^2 \notag \\
    & \left. \qquad - 2(1-2\beta)\left[\ln(\cos^2\rho)-1\right]\Bigg{]}\tan\rho\,\widetilde{\Psi}^{(\lambda)}(\rho)\right\}\widetilde{\Psi}^{(\lambda)}(\rho)\Big{|}_{\rho=\pm\pi/2} = 0\,.
\end{align}
This is satisfied only by the Dirichlet boundary condition $\widetilde{\Psi}^{(\lambda)}(\pm\pi/2)=0$.  Thus, for all values of $\lambda$ the energy fluxes at $\rho=\pm \pi/2$ vanish for some values of $\beta$ if and only if $\widetilde{\mathrm{SL}}(2,\mathbb{R})$-invariant boundary conditions are imposed. 

\subsection{Boundary conditions leading to unitary irreducible representations}\label{CorrIrreps}

In this subsection we discuss the mode functions found in Sec.~\ref{SAC} with reference to the UIRs of the group $\widetilde{\mathrm{SL}}(2,\mathbb{R})$.
Let us start with the cases with $\lambda \geq 3/2$.
In each of these cases the positive-frequency mode functions were found to have spatial components given by Eq.~\eqref{case1sol} and frequencies $\omega_n=\lambda+n$, with $n\in\mathbb{N}\cup\{0\}$. By comparing the ranges for $\lambda$ and the frequencies $\omega=\mu+k$ with the classification given in Sec.~\ref{UIRs}, we find that the eigenvectors of the Casimir operator, \textit{i.e.}\ $\Box$ in Eq.~\eqref{KG}, form a discrete series representation $\mathscr{D}^{+}_{\lambda}$ labeled by the pair $(\lambda,\lambda)$.

Next let us discuss the cases with $1/2 < \lambda < 3/2$. For each of these cases there are four possible $\widetilde{\mathrm{SL}}(2,\mathbb{R})$-invariant self-adjoint extensions of the operator $A$.  Correspondingly, there are four sets of positive-frequency solutions to the Klein-Gordon equation.
The positive-frequency mode functions with the generalized Dirichlet boundary condition, $\phi^{\mathrm{I}}_n(t,\rho)$ in Eq.~\eqref{case1sol}, form the discrete series representation $\mathscr{D}^{+}_{\lambda}$ as in the cases $\lambda \geq 3/2$.

For the positive-frequency mode functions with the generalized Neumann boundary condition, $\phi^{\mathrm{II}}_n(t,\rho)$ in Eq.~\eqref{new-Neumann-modes}, we need to discuss the cases $1/2<\lambda <1$, $\lambda=1$ and $1 < \lambda < 3/2$ separately.  If $1/2<\lambda<1$, then these mode functions form a discrete series representation $\mathscr{D}^{+}_{1-\lambda}$
with the positive frequencies $\omega_{n}=(1-\lambda)+n$, $n\in\mathbb{N}\cup\{0\}$. 

If $\lambda=1$, then the mode function $\phi_0^{\mathrm{II}}(t,\rho)$ has zero frequency. There is no $\widetilde{\mathrm{SL}}(2,\mathbb{R})$-invariant vacuum state because of this mode function.  This situation is analogous to the absence of de~Sitter-invariant vacuum state for the massless minimally-coupled scalar field in de~Sitter space.\cite{allen-dS}
Now, if we let the ladder operator $L_{-}$ act on  $\phi_1^{\mathrm{II}}(t,\rho) \propto \sin\rho\,e^{- it}$ we have by Eq.~\eqref{ladder2}
\begin{align}
    iL_{-}\phi_{1}^{\mathrm{II}}(t,\rho) = \frac{1}{\sqrt{\pi}}\,.
\end{align}
Since a constant function is orthogonal to all mode functions including itself with respect to the Klein-Gordon inner product~\eqref{KG-original-def}, we can identify it with $0$. This amounts to identifying $\phi^{\mathrm{II}}_n(t,\rho)+\textrm{constant}$ with $\phi^{\mathrm{II}}_n(t,\rho)$ for $n\geq 1$.  With this identification, the mode functions $\phi_n^{\mathrm{II}}(t,\rho)$ form the UIR $\mathscr{D}^{+}_1$. Thus, it is possible to define an $\widetilde{\mathrm{SL}}(2,\mathbb{R})$-invariant vacuum state if we ``quotient out'' the zero-frequency sector. (A de~Sitter invariant vacuum state can be constructed also for the massless minimally-couple scalar field in de~Sitter space in this manner. \cite{higuchi-l-inst,kirsten-garriga})

If $1<\lambda<3/2$, then the positive-frequency mode function with $n=1$ satisfies
\begin{align}
    iL_{-}\phi_1^{\mathrm{II}}(t,\rho) = \sqrt{2(\lambda-1)}\,\,\overline{\phi_0^{\mathrm{II}}(t,\rho)}\,,
\end{align}
where Eq.~\eqref{ladder2} has been used.  That is, the infinitesimal transformation of the positive-frequency mode $\phi_1^{\mathrm{II}}(t,\rho)$ leads to a negative-frequency mode $\overline{\phi_0^{\mathrm{II}}(t,\rho)}$.  Thus, the positive-frequency subspace of the solutions to the Klein-Gordon equation and the vacuum state are not invariant under $\widetilde{\mathrm{SL}}(2,\mathbb{R})$ transformations (see Appendix~\ref{App-static-spacetime}).  In this case, the positive-frequency mode functions $\phi_n^{\mathrm{II}}(t,\rho)$, $n\in\mathbb{N}$, together with the negative-frequency mode function $\overline{\phi_0^{\mathrm{II}}(t,\rho)}$ form a representation in the {\it non-unitary discrete series}~\cite{Kitaev} denoted by $\mathscr{F}_{1-\lambda}^{\pm}$.

Next, let us discuss the solutions obeying the mixed boundary conditions for $1/2 < \lambda < 3/2$.  Also in this case, the space of positive-frequency solutions and the vacuum state are non-invariant because positive-frequency mode functions and negative-frequency mode functions mix under the $\widetilde{\mathrm{SL}}(2,\mathbb{R})$ transformations.  This can be seen by using Eq.~\eqref{ladder2} as
\begin{subequations}
\begin{align}
    iL_{-}\phi_0^{\mathrm{III}}(t,\rho) & = \left(\frac{1}{2}-\lambda\right)\overline{\phi_0^{\mathrm{III}}(t,\rho)}\,,\label{L-minus-for-phi0}\\
    iL_{-}\phi_0^{\mathrm{IV}}(t,\rho) & = -\left(\frac{1}{2}-\lambda\right)\overline{\phi_0^{\mathrm{IV}}(t,\rho)}\,.
\end{align}
\end{subequations}
For either of the mixed boundary conditions, the positive- and negative-frequency mode functions together form a non-unitary representation of $\widetilde{\mathrm{SL}}(2,\mathbb{R})$.

For $\lambda=1/2$, the positive-frequency mode functions with the generalized Dirichlet boundary conditions, $\phi_n^{\mathrm{V}}(t,\rho)$, $n\in \mathbb{N}\cup \{0\}$, form a unitary representation because $L_{-}\phi_0^{\mathrm{V}}(t,\rho)=0$ in this case.  This representation is $\mathscr{D}_{1/2}^{+}$, which is a representation in the mock-discrete series, labeled by the pair $(1/2,1/2)$.

We summarize the $\widetilde{\mathrm{SL}}(2,\mathbb{R})$-invariant boundary conditions for each value of $\lambda$ and the properties of the representations formed by the solutions to the Klein-Gordon equation in $\mathrm{AdS}_2$
in Table~\ref{tab:table2}. 

\begin{table}[h!]
  \begin{center}
    \resizebox{.95\textwidth}{!}{
    \begin{minipage}{\textwidth}
     \hspace*{-.5cm}\begin{tabular}{c|l|c|c|c|c}
    \hline
      \textbf{$\boldsymbol{\lambda}$} & \textbf{Spatial mode functions} & $\boldsymbol{\omega}$ & \textbf{Boundary Conditions} & \textbf{Irrep} & \textbf{Unitarity}\\
      \hline\xrowht[()]{20pt}
      $\lambda\geq \frac{3}{2}$ & $(\cos\rho)^{\lambda}P_{n}^{(a,a)}(\sin\rho)$ & $\lambda+n$ & Square integrability & $\mathscr{D}_{\lambda}^{+}$ & Yes \\
      \hline\xrowht[()]{20pt}
      \multirow{4}{*}{$1<\lambda<\frac{3}{2}$} & $(\cos\rho)^{\lambda}P_{n}^{(a,a)}(\sin\rho)$ & $\lambda+n$ & Dirichlet & $\mathscr{D}_{\lambda}^{+}$ & Yes \\ \xrowht[()]{20pt}
      & $(\cos\rho)^{1-\lambda}P_{n}^{(-a,-a)}(\sin\rho)$ & $1-\lambda+n$ & Neumann & $\mathscr{F}_{1-\lambda}^{+}$ & No \\ \xrowht[()]{20pt}
      & $(\cos\rho)^{\lambda}(1-\sin\rho)^{-a}P_{n}^{(-a,a)}(\sin\rho)$ & \multirow{2}{*}{$n+\frac{1}{2}$} & Mixed $\left(\theta=\pi/2\right)$  & -- & No \\ \xrowht[()]{20pt}
      & $(\cos\rho)^{\lambda}(1+\sin\rho)^{-a}P_{n}^{(a,-a)}(\sin\rho)$ & & Mixed $\left(\theta=0\right)$ & -- & No \\
      \hline \xrowht[()]{20pt}
      {$\lambda=1$} &$\cos\,n\rho$, \hspace{.1cm} $n$ odd;\hspace{.1cm} $\sin\,n\rho$,\hspace{.1cm} $n$ even & \multirow{3}{*}
      {$n\,(\geq 1)$} & Dirichlet & \multirow{3}{*}
      {$\mathscr{D}_1^{+}$} & \multirow{3}{*}
      {Yes} \\ \xrowht[()]{20pt}
      & $\sin\, n\rho$, \hspace{.1cm} $n$ odd;\hspace{.1cm} $\cos\,n\rho$,\hspace{.1cm} $n$ even &  & Neumann &  & \\
      \hline \xrowht[()]{20pt}
       \multirow{4}{*}{$\frac{1}{2}<\lambda<1$} & $(\cos\rho)^{\lambda}P_{n}^{(a,a)}(\sin\rho)$ & $\lambda+n$ & Dirichlet & $\mathscr{D}_{\lambda}^{+}$ & Yes \\ \xrowht[()]{20pt}
      & $(\cos\rho)^{1-\lambda}P_{n}^{(-a,-a)}(\sin\rho)$ & $1-\lambda+n$ & Neumann & $\mathscr{D}_{\lambda}^{+}$ & Yes \\ \xrowht[()]{20pt}
      & $(\cos\rho)^{\lambda}(1-\sin\rho)^{-a}P_{n}^{(-a,a)}(\sin\rho)$ & \multirow{2}{*}{$n+\frac{1}{2}$} & Mixed $\left(\theta=\pi/2\right)$  & -- & No \\ \xrowht[()]{20pt}
      & $(\cos\rho)^{\lambda}(1+\sin\rho)^{-a}P_{n}^{(a,-a)}(\sin\rho)$ & & Mixed $\left(\theta=0\right)$ & -- & No \\
      $\lambda=\frac{1}{2}$ & $\sqrt{\cos\rho}\,P_{n}(\sin\rho)$ & $n+\frac{1}{2}$ & Dirichlet & $\mathscr{D}_{1/2}^{+}$ & Yes \\
      \hline
      \multicolumn{6}{c}{Here $a:=\lambda-\frac{1}{2}$, $n\in\mathbb{N}\cup\{0\}$, $k\in\mathbb{Z}$}\\
      \hline
    \end{tabular}
     \caption{Classification of the mode functions}
    \label{tab:table2}
    \end{minipage} }
  \end{center}
\end{table}

\section{The $\widetilde{\mathrm{SL}}(2,\mathbb{R})$-invariant theories with non-invariant positive-frequency subspace}\label{sec-bogoliubov}

In Sec.~\ref{CorrIrreps} we found that some boundary conditions lead to an $\widetilde{\mathrm{SL}}(2,\mathbb{R})$-invariant theory with non-invariant vacuum state if $1/2 < \lambda < 3/2$.  The Klein-Gordon inner product~\eqref{KG-original-def} is $\widetilde{\mathrm{SL}}(2,\mathbb{R})$-invariant with any of these boundary conditions.
This implies that the $\widetilde{\mathrm{SL}}(2,\mathbb{R})$ transformations on the quantum field are Bogoliubov transformations, which mix annihilation and creation operators.
Since a theory with any of these boundary conditions is $\widetilde{\mathrm{SL}}(2,\mathbb{R})$ invariant but the vacuum state is not, it must be possible to find a UIR of this group which the vacuum state belongs to.  In this section we identify this representation. 

We start with the cases with the mixed boundary condition $\widetilde{\Psi}^{(\lambda)}(\pi/2)=D\widetilde{\Psi}^{(\lambda)}(-\pi/2)=0$ with $1/2 < \lambda < 3/2$. (The cases with $\widetilde{\Psi}^{(\lambda)}(-\pi/2)=D\widetilde{\Psi}^{(\lambda)}(\pi/2)=0$ are similar.) 
We expand the field operator $\phi(t,\rho)$ as
\begin{equation}
    \phi(t,\rho) = \sum_{n=0}^\infty\left[ a_n \phi_n^{\mathrm{III}}(t,\rho) + a_n^\dagger\overline{\phi_n^{\mathrm{III}}(t,\rho)}\right]\,.
\end{equation}
Then, the conserved quantum charge for the symmetry generated by $L_\pm$ is
\begin{align}
    \hat{L}_\pm & = \frac{1}{2} \langle \phi,L_{\pm}\phi\rangle_{\mathrm{KG}}\notag \\
    & = i\int_{-\pi/2}^{\pi/2}
    \left[ \phi(t,\rho)\frac{\partial\ }{\partial t}
    L_\pm\phi(t,\rho) - \frac{\partial\phi(t,\rho)}{\partial t}L_\pm\phi(t,\rho)\right]\,. \label{L-pm-quantum}
\end{align}
We use the ladder operators for the Jacobi polynomials~\cite{NIST:DLMF}
\begin{subequations}
\begin{align}
    & (2n+\alpha+\beta)(1-x^2)\frac{d\ }{dx}P_n^{(\alpha,\beta)}(x) \notag \\
    & = n(\alpha-\beta - (2n+\alpha+\beta)x)P_n^{(\alpha,\beta)}(x) + 2(n+\alpha)(n+\beta)P_{n-1}^{(\alpha,\beta)}(x)\,,\\
    & (2n+\alpha+\beta+2)(1-x^2)\frac{d\ }{dx}P_n^{(\alpha,\beta)}(x) \notag \\
    & = (n+\alpha+\beta+1)(\alpha-\beta+(2n+\alpha+\beta+2)x)P_n^{(\alpha,\beta)}(x) \notag \\
    & \quad - 2(n+1)(n+\alpha+\beta+1)P_{n+1}^{(\alpha,\beta)}(x)\,,
\end{align}
\end{subequations}
to find $L_+\phi_n^{\mathrm{III}} = - ik_{n}\phi_{n+1}^{\mathrm{III}}$,
$L_{-}\phi_n^{\mathrm{III}} = - i k_{n-1} \phi_{n-1}^{\mathrm{III}}$ ($n\geq 1$),
$L_{-}\overline{\phi_n^{\mathrm{III}}} = ik_{n}\overline{\phi_{n+1}^{\mathrm{III}}}$, $L_{+}\overline{\phi_n^{\mathrm{III}}} = ik_{n-1}\overline{\phi_{n-1}^{\mathrm{III}}}$ $(n\geq 1)$, where
\begin{equation}
    k_n = \left[\left( \lambda + n +\frac{1}{2} \right)\left( n+\frac{3}{2}-\lambda\right)\right]^{1/2}\,.
\end{equation}
On the other hand, we have $L_{-}\phi_0^{\mathrm{III}} = - i(1/2-\lambda)\overline{\phi_0^{\textrm{III}}}$ (see Eq.~\eqref{L-minus-for-phi0}) and $L_+\overline{\phi_0^\mathrm{III}} = i(1/2-\lambda)\phi_0^{\mathrm{III}}$.
By using these formulas and the orthonormality relations
$\langle\phi_m^{\mathrm{III}},\phi_n^{\mathrm{III}}\rangle_{\mathrm{KG}} = - \langle\overline{\phi_m^{\mathrm{III}}},\overline{\phi_n^{\mathrm{III}}}\rangle_{\mathrm{KG}} = \delta_{mn}$ in Eq.~\eqref{L-pm-quantum}, we find
\begin{align}
    i\hat{L}_+ & =  \sum_{n=0}^\infty k_n a_{n+1}^\dagger a_n - \frac{1}{2}\left(\lambda-\frac{1}{2}\right)(a_0^\dagger)^2\,,\\
     i\hat{L}_-& =  \sum_{n=0}^\infty k_n a_{n}^\dagger a_{n+1} - \frac{1}{2}\left(\lambda-\frac{1}{2}\right)(a_0)^2\,.
\end{align}
Then,
\begin{align}\label{normal-ordered-commutator}
    [\hat{L}_+,\hat{L}_{-}]
    & = \sum_{n=1}^\infty k_n(a_n^\dagger a_n - a_{n+1}^\dagger a_{n+1}) + \frac{1}{2}\left(\lambda-\frac{1}{2}\right)^2 (a_0^\dagger a_0 + a_0 a_0^\dagger)\,.
\end{align}
The first term is ambiguous because of the infinite summation. We require that it should annihilate the vacuum state $|0\rangle$, which amounts to manipulating this summation formally, keeping the operators normal-ordered. 
Thus, we obtain
\begin{align}\label{quantum-com}
    [\hat{L}_+,\hat{L}_{-}] = 2\hat{L}_0\,,
\end{align}
where
\begin{align}
    \hat{L}_0 = \sum_{n=0}^\infty \left(n+\frac{1}{2}\right)a_n^\dagger a_n + \frac{1}{4}\left( \lambda-\frac{1}{2}\right)^2\mathbb{I}\,.
\end{align}
(If we substituted the equality 
$a_n^\dagger a_n - a_{n+1}^\dagger a_{n+1} = a_n a_n^\dagger - a_{n+1}a_{n+1}^\dagger$ into Eq.~\eqref{normal-ordered-commutator} and manipulated the summation formally, we would find an infinite constant added to $\hat{L}_0$.)

By comparing the commutation relation~\eqref{quantum-com} with Eq.~\eqref{Lcommut} we conclude that the operator $\hat{L}_0$ should be identified as the time-translation generator, \textit{i.e.}\ the energy operator.  The vacuum state $|0\rangle$ annihilated by all $a_n$, $n=0,1,2,\ldots$, satisfies $\hat{L}_{-}|0\rangle = 0$ and $\hat{L}_0|0\rangle = [(\lambda-1/2)^2/4]|0\rangle$.  Hence, the state $|0\rangle$ belongs to the representation $\mathscr{D}_{(\lambda-1/2)^2/4}^+$.  The other states in this representation can be found by applying $\hat{L}^+$ repeatedly on $|0\rangle$.  

The theory with $1 < \lambda <3/2$ obtained by imposing the Neumann boundary condition can be studied in the same manner.  In these cases we find
$L_+\phi_n^{\mathrm{II}} = - iq_n \phi_{n+1}^{\mathrm{II}}$ ($n\geq 1$),
$L_{-}\phi_n^{\mathrm{II}} = i q_{n-1}\phi_{n-1}^{\mathrm{II}}$ ($n\geq 2$),
$L_{-}\overline{\phi_n^{\mathrm{II}}} = iq_n\overline{\phi_{n+1}^{\mathrm{II}}}$ ($n\geq 1$) and
$L_{+}\overline{\phi_n^{\mathrm{II}}} = - iq_{n-1}\overline{\phi_{n-1}^{\mathrm{II}}}$ ($n\geq 2$), where
\begin{equation}
    q_n = \sqrt{(n+1)(2-2\lambda + n)}\,.
\end{equation}
On the other hand, we have
$L_{+}\overline{\phi_0^{\mathrm{II}}}= i\sqrt{2(\lambda-1)}\phi_1^{\mathrm{II}}$, $L_{-}\phi_1^{\mathrm{II}} = i\sqrt{2(\lambda-1)}\,\overline{\phi_0^{\mathrm{II}}}$,
$L_{-}\phi_0^{\mathrm{II}} = - i \sqrt{2(\lambda-1)}\,\overline{\phi_1^{\mathrm{II}}}$, 
$L_{+}\overline{\phi_1^{\mathrm{II}}} = - i\sqrt{2(\lambda-1)}\phi_0^{\mathrm{II}}$ and
$L_{-}\overline{\phi_0^{\mathrm{II}}} = L_{+}\phi_0^{\mathrm{II}} = 0$.  Then we find
\begin{align}
    i\hat{L}_+ & = \sum_{n=1}^\infty q_n a_{n+1}^\dagger a_n - \sqrt{2(\lambda-1)}a_1^\dagger a_0^\dagger\,,\\
    i\hat{L}_{-} & = \sum_{n=1}^\infty q_n a_n^\dagger a_{n+1} - \sqrt{2(\lambda-1)}a_1 a_0\,.
\end{align}
Then, in the same way as in the cases with the mixed boundary condition, we find
\begin{equation}
    [\hat{L}_+,\hat{L}_-] = 2\hat{L}_0\,,\\
\end{equation}
where
\begin{equation}
    \hat{L}_0 = \sum_{n=0}^\infty \omega^{\mathrm{II}}_n a_n^\dagger a_n + (\lambda-1)\mathbb{I}\,.
\end{equation}
Thus, the vacuum state in this theory belongs to the representation $\mathscr{D}^+_{\lambda-1}$.

\section{Conclusion}
\label{sec-conclusion}

In this paper we studied the minimally-coupled non-interacting scalar field with mass $M^2 = \lambda(\lambda-1)$ in (the universal cover of) two-dimensional  anti-de~Sitter space ($\mathrm{AdS}_2$).  We first determined all possible boundary conditions based on the requirement that the operator $A$, defined by Eq.~\eqref{spatialKG}, corresponding to the square of the frequency of the mode functions should be extended to a self-adjoint operator, following the general theory of Weyl~\cite{Weyl} and von~Neumann.\cite{Neu} We noted that the parameter $\lambda$ has to be real for $A$ to be positive.\cite{ishi3} Since $M^2$ remains unchanged under $\lambda \leftrightarrow 1-\lambda$, we can restrict $\lambda$ to satisfy $\lambda \geq 1/2$. If the eigenvalues $\omega^2$ for the self-adjoint extension $A_U$ of the operator $A$ are positive, then one can define a quantum field theory with a stationary vacuum state following the standard procedure (see Appendix~\ref{App-static-spacetime}).

For $\lambda \geq 3/2$ the self-adjoint extension of the operator $A$ is unique and determined to correspond to the generalized Dirichlet boundary condition. For $1/2 \leq \lambda < 3/2$ the self-adjoint extensions of $A$ are labeled by a $2\times 2$ unitary matrix $\mathcal{U}$, which parametrizes the boundary conditions at the boundaries as in Eq.~\eqref{SABC}.  These boundary conditions are analogous to the self-adjoint boundary conditions for the free quantum particle in a box.\cite{bonneau}

Next, we determined the self-adjoint boundary conditions which are invariant under the action of the group $\widetilde{\mathrm{SL}}(2,\mathbb{R})$, which is the symmetry group of $\mathrm{AdS}_2$.  The generalized Dirichlet boundary condition is invariant for all $\lambda \geq 1/2$. For $1/2 < \lambda < 3/2$, there are additional invariant boundary conditions, which are the generalized Neumann boundary condition and the mixed boundary conditions consisting of the generalized Dirichlet boundary condition at one end and the generalized Neumann boundary condition at the other.  We also noted that 
these boundary conditions are identical to those obtained by requiring the vanishing of the energy flux at each boundary. It will be interesting to investigate deeper connections, if any, between these two requirements.

The set of solutions to the Klein-Gordon equation satisfying an invariant boundary condition forms a representation of the group $\widetilde{\mathrm{SL}}(2,\mathbb{R})$, but this representation may not be unitary.  For the stationary vacuum state to be invariant under the $\widetilde{\mathrm{SL}}(2,\mathbb{R})$ symmetry, the positive-frequency subset of the solutions must form a unitary representation.  We found that the positive-frequency solutions form a unitary representation for the generalized Dirichlet boundary condition for all $\lambda\geq 1/2$ and for the generalized Neumann boundary condition for $1/2 < \lambda < 1$.  The generalized Neumann boundary condition does not lead to a unitary representation for $1 < \lambda < 3/2$, and the mixed boundary conditions do not lead to a unitary representation for any value of $\lambda$.
For $\lambda =1$ ($M^2=0$), the Neumann boundary condition allows spatially constant solutions.  In this case, the spatially non-constant positive-frequency mode functions, $\phi_n^{\mathrm{II}}(t,\rho)$, $n\in \mathbb{N}$, form a UIR.

Finally, we studied the cases where the boundary condition is $\widetilde{\mathrm{SL}}(2,\mathbb{R})$-invariant but the positive-frequency subspace is not, \textit{i.e.}\ the cases with the mixed boundary conditions with $1/2 < \lambda < 3/2$ and those with the Neumann boundary condition with $1 < \lambda < 3/2$ in more detail.  In particular, we found that the vacuum state, which is not invariant under the $\widetilde{\mathrm{SL}}(2,\mathbb{R})$ transformations, belongs to a UIR in each case.

\acknowledgments

We thank Simon Eveson, Chris Fewster and Bernard Kay for useful discussions.  The work of L.\ S. was supported in part by a Studentship under Grant Number
EP/N509802/1 from EPSRC and a departmental Doctoral Prize Studentship, and the work of D.\ S.\ B. was supported by an Overseas Research Studentship from the University of York.


study.

\appendix

\section{Free scalar field in static spacetime}\label{App-static-spacetime}

In this appendix we review non-interacting scalar field theory in static spacetime with a stationary vacuum state.\cite{ashtekar-scalar,kay-scalar,waldQFTCSbook}   Consider a static spacetime with the following metric:
\begin{align}\label{AppA-static}
    ds^2 = -N^2dt^2 + g_{ab}dx^a dx^b\,,
\end{align}
where $N$ and $g_{ab}$ are independent of the time variable $t$.  The Lagrangian for the minimally coupled scalar field with mass $M$ in this spacetime is
\begin{subequations}\label{scalar-lagrangian}
\begin{align}
  L & = \int_\Sigma d\mathbf{x}\,\mathcal{L}\,,\\
  \mathcal{L} & = \frac{1}{2}\sqrt{g}
 \left[ N^{-1}(\partial_t\phi)^2 - Ng^{ab}(\partial_a\phi)(\partial_b\phi) - N M^2\phi^2\right]\,,
\end{align}
\end{subequations}
where $g := \textrm{det}(g_{ab})$.  The spacelike hypersurface $\Sigma$ is that of constant $t$.
The conjugate momentum density is
\begin{align}
    \pi(t,\mathbf{x}) & = \frac{\partial \mathcal{L}}{\partial (\partial_t\phi(t,\mathbf{x}))} \nonumber \\
    & = \frac{\sqrt{g(\mathbf{x})}}{N(\mathbf{x})}\partial_t\phi(t,\mathbf{x})\,.
\end{align}
Hence, the equal-time canonical commutation relations read
\begin{subequations}\label{AppA-et-comm}
\begin{align}
    \left[\phi(t,\mathbf{x}), \partial_t \phi(t,\mathbf{x}')\right] & = i \frac{N(\mathbf{x})}{\sqrt{g(\mathbf{x})}}\delta(\mathbf{x},\mathbf{x}')\,,\\
    \left[\phi(t,\mathbf{x}),\phi(t,\mathbf{x}')\right] & = \left[\partial_t\phi(t,\mathbf{x}),\partial_t\phi(t,\mathbf{x}')\right] = 0\,,
\end{align}
\end{subequations}
where
\begin{align}
    \int_{\Sigma}d\mathbf{x}'\,\delta(\mathbf{x},\mathbf{x}')f(\mathbf{x}') = f(\mathbf{x})\,,
\end{align}
for any smooth compactly supported function $f$ on $\Sigma$.

The Lagrange's equation derived from the Lagrangian~\eqref{scalar-lagrangian} is
\begin{align}
    -\frac{1}{N^2}\frac{\partial^2\phi}{\partial t^2} + \frac{1}{\sqrt{g}\,N}\partial_a(\sqrt{g}\,g^{ab} N\partial_b \phi) - M^2\phi = 0\,.
\end{align}
Solutions to this equation can be found in the form
\begin{align}\label{AppA-phi-def}
    \phi_\sigma(t,\mathbf{x})
    = \frac{1}{\sqrt{2\omega_\sigma}}\Psi_\sigma(\mathbf{x}) e^{-i\omega_\sigma t}\,,
\end{align}
where
\begin{align}
    A\Psi_\sigma = \omega_\sigma^2\Psi_\sigma\,,
\end{align}
with the differential operator $A$ defined by
\begin{align}
  A :=  - \frac{N}{\sqrt{g}}\partial_a\sqrt{g}\,g^{ab} N\partial_b + M^2N^2\,.
\end{align}
Note that the operator $A$ satisfies
\begin{align}\label{AppA-symmetry}
    \langle\Psi_1,A\Psi_2\rangle = \langle A\Psi_1,\Psi_2\rangle\,,
\end{align}
where
\begin{align}\label{AppA-inner-product}
    \langle\Psi_1,\Psi_2\rangle : = \int_\Sigma d\mathbf{x}\,\frac{\sqrt{g}}{N}
    \overline{\Psi_1}\Psi_2\,,
\end{align}
if the boundary terms vanish.

Suppose the operator $A$ with an appropriate domain is self-adjoint with the inner product~\eqref{AppA-inner-product}, \textit{i.e.}\ that Eq.~\eqref{AppA-symmetry} is satisfied and that the domain of the adjoint $A^\dagger$ equals the domain of $A$ itself.  Suppose further that the spectrum of the operator $A$ is discrete and $\omega_\sigma^2 > 0$ for all $\sigma$.  Then, the eigenfunctions $\Psi_\sigma$ of the operator $A$ are complete, and the quantum field $\phi(t,\mathbf{x})$ can be expanded as
\begin{align}\label{AppA-expansion}
    \phi(t,\mathbf{x}) = \sum_\sigma\left[ a_\sigma \phi_\sigma(t,\mathbf{x}) + a_\sigma^\dagger \overline{\phi_\sigma(t,\mathbf{x})}\right]\,,
\end{align}
with $\omega_\sigma > 0$, where the mode functions $\phi_\sigma(t,\mathbf{x})$ are defined by Eq.~\eqref{AppA-phi-def}.

Note that, if $\omega_\sigma\neq\omega_{\sigma'}$, then the functions $\Psi_\sigma$ and $\Psi_{\sigma'}$ are orthogonal because of the relation~\eqref{AppA-symmetry}.
This allows us to normalize the functions $\Psi_\sigma$ as
\begin{align}\label{AppA-normalisation}
    \langle\Psi_\sigma,\Psi_{\sigma'}\rangle =\delta_{\sigma\sigma'}\,.
\end{align}
The completeness of the functions $\Psi_\sigma$ and the normalization condition~\eqref{AppA-normalisation} imply
\begin{align}
    \sum_\sigma \Psi_\sigma(\mathbf{x})\overline{\Psi_\sigma(\mathbf{x}')} = \frac{N(\mathbf{x})}{\sqrt{g(\mathbf{x})}}\delta(\mathbf{x},\mathbf{x}')\,.
\end{align}
This completeness relation allows one to show that the equal-time commutation relations~\eqref{AppA-et-comm} are equivalent to the commutation relations among the annihilation and creation operators:
\begin{align}
    [a_\sigma,a^\dagger_{\sigma'}] = \delta_{\sigma\sigma'}\,,
\end{align}
with all other commutators among $a_\sigma$ and $a_{\sigma'}^\dagger$ vanishing.  The vacuum state $|0\rangle$ is defined by the requirement that 
\begin{align}\label{AppA-vacuum}
a_\sigma|0\rangle = 0\,,   
\end{align} 
for all $\sigma$.

Now, let $\Lambda$ be a Killing vector of the spacetime and let
\begin{align}
    \pounds_\Lambda \phi_\sigma(t,\mathbf{x}) = \sum_{\sigma'}\Lambda_{\sigma\sigma'}\phi_{\sigma'}(t,\mathbf{x}) + \sum_{\sigma'}\tilde{\Lambda}_{\sigma\sigma'}\overline{\phi_{\sigma'}(t,\mathbf{x})}\,,
\end{align}
where $\pounds_\Lambda$ denotes the Lie derivative with respect to $\Lambda$. Substituting this expression into Eq.~\eqref{AppA-expansion}, we have
\begin{align}
    \pounds_\Lambda \phi(t,\mathbf{x}) = \sum_\sigma\sum_{\sigma'}\left[(a_{\sigma'}\Lambda_{\sigma'\sigma}+a_{\sigma'}^\dagger\overline{\tilde{\Lambda}_{\sigma'\sigma}})\phi_\sigma(t,\mathbf{x}) + (a_{\sigma'}^\dagger\overline{\Lambda_{\sigma'\sigma}}+a_{\sigma'}\tilde{\Lambda}_{\sigma'\sigma})\overline{\phi_\sigma(t,\mathbf{x})}\right]\,. 
\end{align}
Thus, the infinitesimal transformation of the annihilation operators $a_\sigma$ corresponding to the symmetry transformation generated by $\Lambda$ is given by
\begin{align}
    \delta_\Lambda a_\sigma = \sum_{\sigma'}\left(a_{\sigma'}\Lambda_{\sigma'\sigma}+ a_{\sigma'}^\dagger \overline{\tilde{\Lambda}_{\sigma'\sigma}}\right)\,.
\end{align}
Hence, for the vacuum state $|0\rangle$ defined by Eq.~\eqref{AppA-vacuum} to be invariant under the spacetime symmetry transformation corresponding to the Killing vector $\Lambda$, we need to have $\tilde{\Lambda}_{\sigma\sigma'} = 0$.  That is,
\begin{align}
    \pounds_\Lambda \phi_\sigma(t,\mathbf{x}) = \sum_{\sigma'}\Lambda_{\sigma\sigma'}\phi_{\sigma'}(t,\mathbf{x})\,.
\end{align}
Thus, for the vacuum state $|0\rangle$ to be invariant under this symmetry transformation, the positive-frequency solutions $\phi_\sigma(t,\mathbf{x})$ must transform among themselves without any component of negative-frequency solutions.  Note that $|0\rangle$ is stationary, \textit{i.e.}\ invariant under time-translation symmetry $T$ with $\pounds_T\phi_\sigma(t,\mathbf{x}) = -i\omega_\sigma\phi_\sigma(t,\mathbf{x})$. 
\section{The operator $A$ with $M^2 < -1/4$}\label{App-negative-definite}

In this appendix we demonstrate that the operator $A$ with $M^2 < -1/4$ is unbounded from below.  In this case we have
\begin{equation}
    A = - \frac{d^2\ }{d\rho^2} - \frac{1/4 + a}{\cos^2\rho}\,,
\end{equation}
with $a > 0$.  We first observe
\begin{align}\label{eta-eta}
    \int_{-\eta}^{\eta} (\cos\rho)^{1/2}A(\cos\rho)^{1/2}d\rho &  = -2a\ln(\sec\eta+\tan\eta) + \frac{1}{2}\sin\eta\,, 
\end{align}
for $0 < \eta < \pi/2$. Notice that this integral diverges to $-\infty$ as $\eta\to\pi/2$. Let $\pi/6 < \eta < \pi/2$ and  $\epsilon = (\pi/2-\eta)/2$.  Then $0 < \eta - \epsilon < \eta+\epsilon < \pi/2$.

Let $f\in \dom{A}$ be defined by
\begin{equation}
    f(\rho) : = \begin{cases} (\cos\rho)^{1/2} & \textrm{if}\ \ |\rho| \leq \eta - \epsilon\,, \\
    (\cos\rho)^{1/2}\chi((|\rho|-\eta)/\epsilon) & \textrm{if}\ \  |\rho| \geq\eta - \epsilon\,, \end{cases}
\end{equation}
where $\chi$ is a smooth monotonically-decreasing function satisfying the condition 
that $\chi(x) = 1$ if $x \leq -1$ and $\chi(x)=0$ if $x \geq 1$.  We have $f\in \dom{A}$ because $f(\rho)=0$ if $\eta+\epsilon \leq |\rho| < \pi/2$. 
We have
\begin{equation}
    \int_{-\pi/2}^{\pi/2}|f(\rho)|^2 d\rho \leq 2\,.
\end{equation}
and
\begin{equation}\label{need-bounded}
    \int_{-\pi/2}^{\pi/2} \overline{f(\rho)}Af(\rho)d\rho = \int_{-\eta + \epsilon}^{\eta-\epsilon} \overline{f(\rho)}Af(\rho)d\rho + 2\int_{\eta-\epsilon}^{\eta+\epsilon} \overline{f(\rho)}Af(\rho)d\rho\,.
\end{equation}
Since the first integral diverges to $-\infty$ as $\eta\to \pi/2$ by Eq.~\eqref{eta-eta}, if the second integral is bounded in this limit, then the operator $A$ is unbounded from below.  

For $\eta - \epsilon < \rho < \eta+\epsilon$ we find
\begin{align}
    f(\rho)Af(\rho) & = \chi((\rho-\eta)/\epsilon)\left[ - \frac{1}{\epsilon^2}\cos\rho\chi''( (\rho-\eta)/\epsilon) + \frac{1}{\epsilon}
    \sin\rho\chi'((\rho-\eta)/\epsilon)\right. \notag \\
    & \quad + \left. \left( - \frac{a}{\cos\rho} + \frac{1}{4}\cos\rho\right)\chi((\rho-\eta)/\epsilon)\right]\,.
\end{align}
Let $|\chi''(x)|\leq C_2$, $|\chi'(x)|\leq C_1$ and recall $|\chi(x)|\leq 1$.  Then,
\begin{align}
    |f(\rho)Af(\rho)| & \leq \frac{C_2}{\epsilon^2}\sin 3\epsilon
    + \frac{C_1}{\epsilon} +  \frac{a}{\sin \epsilon} + \frac{1}{4}\,.
\end{align}
Then,
\begin{align}
\left|    \int_{\eta-\epsilon}^{\eta+\epsilon}\overline{f(\rho)}Af(\rho)d\rho\right| 
& \leq  \int_{\eta-\epsilon}^{\eta+\epsilon}|\overline{f(\rho)}Af(\rho)|d\rho \notag\\ 
& \leq \frac{2C_2}{\epsilon}\sin 3\epsilon + 2C_1 + \frac{2a\epsilon}{\sin\epsilon} + \frac{\epsilon}{2}\notag \\
& \to  6C_2 + 2C_1 + 2a\,,
\end{align}
as $\epsilon = (\pi/2-\eta)/2 \to 0$.  Hence, the second term in Eq.~\eqref{need-bounded} is indeed bounded and the operator $A$ is unbounded from below.

\section{The closure of the operator $A$}
\label{App-A-bar-boundary}

In this appendix we demonstrate that, if $\Psi \in \dom{\bar{A}}$, then $\widetilde{\Psi}^{(\lambda)}(\pm\pi/2) = D\widetilde{\Psi}^{(\lambda)}(\pm\pi/2) = 0$, where $\bar{A}$ denotes the closure of the operator $A$ and where $\widetilde{\Psi}^{(\lambda)}$ and $D\widetilde{\Psi}^{(\lambda)}$ are defined by Eqs.~\eqref{weightedsol} and \eqref{weightedsol2}.
First we examine the case with $1/2 < \lambda < 3/2$.  Let $\Phi_1(\rho)$ and $\Phi_2(\rho)$ be smooth
functions whose support is in $[0,\pi/2]$ and which take the values $(\cos\rho)^\lambda$ and $(\cos\rho)^{1-\lambda}$,
respectively, for $\rho\in [\pi/4,\pi/2)$.  Then since $\lambda$ and $1-\lambda$ are both larger 
than $-1/2$ so that $\Phi_1,\Phi_2 \in L^2[-\pi/2,\pi/2]$
 and since, for $\rho \in [\pi/4,\pi/2)$,
\begin{subequations}
\begin{align}
\left( - \frac{d^2\ }{d\rho^2} + \frac{\lambda(\lambda-1)}{\cos^2\rho}\right) \Phi_1(\rho)
& =  \lambda^2(\cos\rho)^\lambda,\\
\left( - \frac{d^2\ }{d\rho^2} + \frac{\lambda(\lambda-1)}{\cos^2\rho}\right) \Phi_2(\rho)
& =  (1-\lambda)^2(\cos\rho)^{1-\lambda}\,,
\end{align}
\end{subequations}
we have $A^\dagger\Phi_1, A^\dagger\Phi_2 \in L^2[-\pi/2,\pi/2]$ and therefore $\Phi_1,\Phi_2 \in \dom{A^\dagger}$.  

Now, suppose $\Psi \in \dom{\bar{A}}$.  Since $\bar{A}=(A^\dagger)^\dagger$ (see Reed and Simon~\cite{reed-one}), we have by definition
\begin{align}
\langle c_1 \Phi_1+c_2\Phi_2, \bar{A}\Psi\rangle - \langle c_1 A^\dagger\Phi_1+c_2A^\dagger\Phi_2,\Psi\rangle = 0\,,
\end{align}
where $c_1,c_2\in\mathbb{C}$.
This can be written as
\begin{align}
\lim_{a\to \pi/2}\int_0^a 
\left\{ \frac{d^2\ }{d\rho^2}\left[\overline{c_1} \Phi_1(\rho) + \overline{c_2} \Phi_2(\rho)\right]\Psi(\rho)
- \left[\overline{c_1}\Phi_1(\rho) + \overline{c_2}\Phi_2(\rho)\right]\frac{d^2\ }{d\rho^2}\Psi(\rho)\right\}d\rho = 0.
\end{align}
Then, by integration by parts we have
\begin{align}
& \lim_{\rho\to\pi/2}\left\{
(1-2\lambda)\overline{c_1}(\cos\rho)^{\lambda-1}\sin\rho \Psi(\rho) \right. \notag\\
& \left.
\qquad - \left[ \overline{c_1}(\cos\rho)^{2\lambda-1} + \overline{c_2}\right](\cos\rho)^{2-2\lambda}\frac{d\ }{d\rho}
\left[ (\cos\rho)^{\lambda-1}\Psi(\rho)\right] \right\} = 0. \label{general-case}
\end{align}
Let $c_1=0$ and $c_2=1$. Then,
\begin{align}
\lim_{\rho\to\pi/2}(\cos\rho)^{2-2\lambda}\frac{d\ }{d\rho}
\left[ (\cos\rho)^{\lambda-1}\Psi(\rho)\right] = 0.
\end{align}
That is,
$D\widetilde{\Psi}^{(\lambda)}(\pi/2) = 0$.
Next, let $c_1=1$ and $c_2=0$.  Then since $(\cos\rho)^{2\lambda-1} \to 0$ and $\sin\rho\to 1$ as 
$\rho \to \pi/2$, we find
\begin{align}
\lim_{\rho\to \pi/2}(\cos\rho)^{\lambda-1}\Psi(\rho) = 0.
\end{align}
That is,
$\widetilde{\Psi}^{(\lambda)}(\pi/2) = 0$.
We can construct a similar argument to show that
$D\widetilde{\Psi}^{(\lambda)}(-\pi/2) = \widetilde{\Psi}^{(\lambda)}(-\pi/2) = 0$.

For $\lambda=1/2$ we can let $\Phi_1(\rho)=(\cos\rho)^{1/2}$ and $\Phi_2(\rho) = (\cos\rho)^{1/2}[\ln(\cos^2\rho) -1]$ for $\rho\in [\pi/4,\pi/2)$ and let them vanish for $\rho \in [-\pi/2,0]$.  We find that
$\Phi_2$ is also in $\dom{A^\dagger}$ because
\begin{align}
\left( - \frac{d^2\ }{d\rho^2} - \frac{1}{4\cos^2\rho}\right)
(\cos\rho)^{1/2}\left[\ln(\cos^2\rho)-1\right] = & \frac{1}{4}(\cos\rho)^{1/2}\left[\ln(\cos^2\rho)-1\right]\nonumber \\
& + 2(\cos\rho)^{1/2}.
\end{align}
Proceeding in the same way as before,
if $\Psi\in \dom{\bar{A}}$, then we find, instead of Eq.~(\ref{general-case}), 
\begin{align}
& \lim_{\rho\to \pi/2}\left\{ \frac{2\overline{c_1}\sin\rho}{(\cos\rho)^{1/2}\left[\ln(\cos^2\rho)-1\right]}\Psi(\rho) \right.
\nonumber \\
& \left. - \left[ \frac{\overline{c_1}}{\left[\ln(\cos^2\rho)-1\right]}
+ \overline{c_2}\right](\cos\rho)\left[\ln(\cos^2\rho)-1\right]^2\frac{d\ }{d\rho}
\left(\frac{\Psi(\rho)}{(\cos\rho)^{1/2}\left[\ln(\cos^2\rho)-1\right]}\right)\right\} \nonumber \\
& = 0.
\end{align}
By choosing $c_1=0$ and $c_2=1$, we find
\begin{align}
\lim_{\rho\to \pi/2}(\cos\rho)\left[\ln(\cos^2\rho)-1\right]^2\frac{d\ }{d\rho}
\left(\frac{\Psi(\rho)}{(\cos\rho)^{1/2}\left[\ln(\cos^2\rho)-1\right]}\right)
 = 0.
\end{align}
That is,
$D\widetilde{\Psi}^{(1/2)}(\pi/2) = 0$.
Next we choose $c_1=1$ and $c_2=0$ and we find, since $\sin\rho\to 1$ and $\ln(\cos^2\rho) \to -\infty$ as
$\rho\to \pi/2$,
\begin{align}
 \lim_{\rho\to \pi/2} \frac{\Psi(\rho)}{(\cos\rho)^{1/2}\left[\ln(\cos^2\rho)-1\right]}=0.
\end{align}
That is,
$\widetilde{\Psi}^{(1/2)}(\pi/2) = 0$.
We can argue in a similar manner to conclude
$D\widetilde{\Psi}^{(1/2)}(-\pi/2) =
\widetilde{\Psi}^{(1/2)}(-\pi/2)  =  0$.
\
In fact, it is possible to show that if $\Psi \in \dom{\bar{A}}$, then
\begin{equation}\label{stronger-limit}
\lim_{\rho \to \pm \pi/2} (\cos\rho)^{-3/2}\Psi(\rho) = 0\,,
\end{equation}
if $1/2\leq \lambda < 3/2$, which is stronger than one of the results, $\widetilde{\Psi}^{(\lambda)}(\pm \pi/2) = 0$.

\section{Relation between the two descriptions of self-adjoint extensions}\label{appendix}

In this appendix we find a one-to-one map between the unitary matrix $U_{\mathrm{M}}$ characterizing the operator $U$ defined by Eq.~\eqref{Uaction} and the unitary matrix $\mathcal{U}$ defined by Eq.~\eqref{first-self-adjoint-BC}. These two matrices characterize the self-adjoint extensions of the operator $A$ in two different ways. We write the matrix $U_\mathrm{M}$ as $U$ in this appendix for simplicity.

Let $\Psi\in \dom{A_U}$, where $A_U$ is a self-adjoint extension of $A$, and let
$\Phi\in\mathscr{S}\subset \mathscr{K}_+\oplus\mathscr{K}_{-}$. (Recall that $\dom{A_U} = \dom{\bar{A}} \oplus \mathscr{S}$.)
Then, from Eqs.~\eqref{Uaction} and \eqref{DomExtA} we find
\begin{align}
\Phi=\begin{pmatrix}
u_{11}\overline{\Phi^{(1)}}+u_{12}\overline{\Phi^{(2)}} \\[.5em]
u_{21}\overline{\Phi^{(1)}}+u_{22}\overline{\Phi^{(2)}} 
\end{pmatrix}\,,
\end{align}
with $\Phi^{(1)},\Phi^{(2)}\in \mathscr{K}_{+}$.
Since the operator $A_U$ is symmetric, it follows that
\begin{align}\label{original-condition}
\inpr{\Psi}{A_U\Phi}-\inpr{A_U\Psi}{\Phi}=0\,.
\end{align}
By integration by parts, this equation can be written in terms of the boundary values as
\begin{align}\label{Ap2}
\left.\Psi\left(\rho\right)\overline{G'_{j}\left(\rho\right)}\right|_{\pi/2}-\left.\Psi'\left(\rho\right)\overline{G_{j}\left(\rho\right)}\right|_{\pi/2}-\left.\Psi\left(\rho\right)\overline{G'_{j}\left(\rho\right)}\right|_{-\pi/2}+\left.\Psi'\left(\rho\right)\overline{G_{j}\left(\rho\right)}\right|_{-\pi/2}=0\,,
\end{align}
where we have defined
\begin{align}
G_{j}(\rho)=u_{j1}\overline{\Phi^{(1)}(\rho)}+u_{j2}\overline{\Phi^{(2)}(\rho)}\,,\hspace{.5cm} j=1,2\,.
\end{align}

From Eq.~\eqref{defSolutions1} we know that the solutions in the deficiency spaces $\mathscr{K}_+\oplus\mathscr{K}_{-}$ behave at the boundaries in a way similar to the solutions of the original eigenvalue problem.  
Thus, we can write Eq.~\eqref{Ap2} as follows:
\begin{align}\label{Ap2-modified}
  &  \widetilde{\Psi}^{(\lambda)}\left(\pi/2\right)\overline{D\widetilde{G}_j^{(\lambda)}\left(\pi/2\right)}-D\widetilde{\Psi}^{(\lambda)}\left(\pi/2\right)\overline{\widetilde{G}_j^{(\lambda)}\left(\pi/2\right)} \notag \\
& \qquad    -\widetilde{\Psi}^{(\lambda)}\left(-\pi/2\right)\overline{D\widetilde{G}_j^{(\lambda)}\left(-\pi/2\right)}+D\widetilde{\Psi}^{(\lambda)}\left(-\pi/2\right)\overline{\widetilde{G}_j^{(\lambda)}\left(-\pi/2\right)} = 0\,,
\end{align}
where $\widetilde{G}_j^{(\lambda)}(\rho)$ and $D\widetilde{G}_j^{(\lambda)}(\rho)$ are defined similarly to $\widetilde{\Psi}^{(\lambda)}(\rho)$ and $D\widetilde{\Psi}^{(\lambda)}(\rho)$.
Next, we define
\begin{subequations}
\begin{align}
    \mathcal{A} & := \begin{pmatrix} D\widetilde{\Phi}^{(1,\lambda)}\left(\pi/2\right) & 0 \\ 0 & D\widetilde{\Phi}^{(2,\lambda)}\left(\pi/2\right)\end{pmatrix} = \begin{pmatrix} - D\widetilde{\Phi}^{(1,\lambda)}\left(-\pi/2\right) & 0 \\ 0 & D\widetilde{\Phi}^{(2,\lambda)}\left(-\pi/2\right)\end{pmatrix}\,,
    \label{mathcalA}\\
    \mathcal{B} & :=\begin{pmatrix} \widetilde{\Phi}^{(1,\lambda)}\left(\pi/2\right) & 0 \\ 0 & \widetilde{\Phi}^{(2,\lambda)}\left(\pi/2\right)\end{pmatrix} = \begin{pmatrix}  \widetilde{\Phi}^{(1,\lambda)}\left(-\pi/2\right) & 0 \\ 0 & - \widetilde{\Phi}^{(2,\lambda)}\left(-\pi/2\right) \end{pmatrix}\,, \label{mathcalB}
\end{align}
\end{subequations}
where $\widetilde{\Phi}^{(j,\lambda)}(\rho)$ and $D\widetilde{\Phi}^{(j,\lambda)}(\rho)$ are defined from $\Phi^{(j)}$, $j=1,2$, in the same way as $\widetilde{\Psi}^{(\lambda)}$ and $D\widetilde{\Psi}^{(\lambda)}$ are defined from $\Psi$.
The second equalities in Eqs.~\eqref{mathcalA} and \eqref{mathcalB} follow from the fact that $\Phi^{(1)}$ and $\Phi^{(2)}$ are even and odd, respectively.  Then, Eq.~\eqref{Ap2-modified} can be written in a matrix form as
\begin{align}\label{matrixbc1}
\left(\overline{\mathcal{A}}+\overline{U}\mathcal{A}\right)\overrightarrow{\Psi}
=\left(\overline{\mathcal{B}}+\overline{U}\mathcal{B}\right)\overrightarrow{D\Psi}\,,
\end{align}
where
\begin{align}\label{vecpsi}
\overrightarrow{\Psi}:=\begin{pmatrix}
\widetilde{\Psi}^{(\lambda)}\left(\pi/2\right)+\widetilde{\Psi}^{(\lambda)}\left(-\pi/2\right)\\[.5em]
\widetilde{\Psi}^{(\lambda)}\left(\pi/2\right)-\widetilde{\Psi}^{(\lambda)}\left(-\pi/2\right)\end{pmatrix}\,,\hspace{.5cm}\overrightarrow{D\Psi}:=\begin{pmatrix}
D\widetilde{\Psi}^{(\lambda)}\left(\pi/2\right)-D\widetilde{\Psi}^{(\lambda)}\left(-\pi/2\right)\\[.5em]
D\widetilde{\Psi}^{(\lambda)}\left(\pi/2\right)+D\widetilde{\Psi}^{(\lambda)}\left(-\pi/2\right)
\end{pmatrix}\,.
\end{align}

It is useful to note that, by expressing the relation $\langle\Phi^{(j)},A_U\Phi^{(j)}\rangle -\langle A_U\Phi^{(j)},\Phi^{(j)}\rangle = 4i$, $j=1,2$, in terms of the boundary values $\widetilde{\Phi}^{(j,\lambda)}(\pm\pi/2)$ and $D\widetilde{\Phi}^{(j,\lambda)}(\pm\pi/2)$, one finds
\begin{align}\label{useful-relation}
 \mathcal{B}\overline{\mathcal{A}} - \mathcal{A}\overline{\mathcal{B}} = 2i\mathbb{I}\,.
\end{align}
We rearrange Eq.~\eqref{matrixbc1} as
\begin{align}\label{matrixbc2}
\left[\overline{\mathcal{B}}-i\overline{\mathcal{A}}+\overline{U}(\mathcal{B}-i\mathcal{A})\right]\left(\overrightarrow{D\Psi}-i\overrightarrow{\Psi}\right)=-\left[\overline{\mathcal{B}}+i\overline{\mathcal{A}}+\overline{U}(\mathcal{B}+i\mathcal{A})\right]\left(\overrightarrow{D\Psi}+i\overrightarrow{\Psi}\right)\,.
\end{align}
The matrices $\overline{\mathcal{B}}\pm i\overline{\mathcal{A}}+\overline{U}(\mathcal{B}\pm i\mathcal{A})$ are invertible because the relation~\eqref{useful-relation} implies that there are no non-trivial solutions $\vec{a}$ to either of the equations $\|(\overline{\mathcal{B}}\pm i\overline{\mathcal{A}})\vec{a}\|^2 = \|(\mathcal{B} \pm i\mathcal{A})\vec{a}\|^2$. 
Then, the matrix $\widetilde{\mathcal{U}}$ defined by
\begin{align}\label{Ucorresp}
\widetilde{\mathcal{U}}:=-\left[\overline{\mathcal{B}}-i\overline{\mathcal{A}}+\overline{U}(\mathcal{B}-i\mathcal{A})\right]^{-1}\left[\overline{\mathcal{B}}+i\overline{\mathcal{A}}+\overline{U}(\mathcal{B}+i\mathcal{A})\right]\,,
\end{align}
is unitary and the map $U\mapsto \widetilde{\mathcal{U}}$ is a bijection~\cite{HigSer} as we show below. Thus, the self-adjoint extensions characterized by the unitary matrix $U$ is indeed equivalently characterized by another unitary matrix $\widetilde{\mathcal{U}}$ which specifies the boundary conditions.  

The unitarity of $\widetilde{\mathcal{U}}$ follows from
\begin{align}\label{unitarity-proof}
V_1V_1^\dagger - V_2 V_2^\dagger = 4 (\mathbb{I} - \overline{U}\,\overline{U}^\dagger)\,,
\end{align}
where
\begin{subequations}
\begin{align}
V_1 & : = \overline{\mathcal{B}}-i\overline{\mathcal{A}}+\overline{U}(\mathcal{B}-i\mathcal{A})\,,\\
V_2 & := \overline{\mathcal{B}}+i\overline{\mathcal{A}}+\overline{U}(\mathcal{B}+i\mathcal{A})\,,
\end{align}
\end{subequations}
since $\widetilde{\mathcal{U}} = - V_1^{-1}V_2$ and $\overline{U}\,\overline{U}^\dagger=\mathbb{I}$. Equation~\eqref{unitarity-proof} results from Eq.~\eqref{useful-relation}.
Next, we show that the map $U\mapsto \widetilde{\mathcal{U}}$ is a bijection by demonstrating that the matrix $U$ satisfying $V_1\widetilde{\mathcal{U}}=-V_2$ for a given unitary matrix $\widetilde{\mathcal{U}}$ exists and is unique.  This equation is solved uniquely for $\overline{U}$ if and only if the homogeneous equation 
\begin{align}\label{prelim-eq}
    \overline{U}(\mathcal{B}-i\mathcal{A})\widetilde{\mathcal{U}} = - \overline{U}(\mathcal{B}+i\mathcal{A})\,,
\end{align}
admits only the trivial solution $\overline{U}=0$. Indeed, if Eq.~\eqref{prelim-eq} is satisfied, then
\begin{align}
 \overline{U}(\mathcal{B}-i\mathcal{A})(\overline{\mathcal{B}}+i\overline{\mathcal{A}})\overline{U}^\dagger = \overline{U}(\mathcal{B}+i\mathcal{A})(\overline{\mathcal{B}}-i\overline{\mathcal{A}})\overline{U}^\dagger\,.
\end{align}
Then, by Eq.~\eqref{useful-relation} we find $\overline{U}\,\overline{U}^\dagger = 0$, which implies $\overline{U}=0$.

Thus, we can write Eq.~\eqref{matrixbc2} as
\begin{align}
(\mathbb{I}-\widetilde{\mathcal{U}})\overrightarrow{D\Psi}=i(\mathbb{I}+\widetilde{\mathcal{U}})\overrightarrow{\Psi}\,,
\end{align}
where $\widetilde{\mathcal{U}}$ is unitary.
Then, by defining
\begin{align}
    \mathcal{U}:= \frac{1}{2}\begin{pmatrix} 1 & 1 \\ -1 & 1 \end{pmatrix} \widetilde{\mathcal{U}}\begin{pmatrix} 1 & - 1 \\ 1 & 1 \end{pmatrix}\,,
\end{align}
we arrive at Eq.~\eqref{SABC}.
It is clear that the map $U \mapsto \mathcal{U}$ is a bijection because the map $U \mapsto \mathcal{\widetilde{U}}$ is.

\section{Boundary conditions with negative eigenvalues of $A_U$}
\label{App-negative-eigenvalues}

In this example we let $\lambda = 1$ so that the eigenvalue problem is given by
\begin{align}\label{AppD-eigenvalue-eq}
    - \frac{d^2\ }{d\rho^2}\Psi(\rho) = \omega^2\Psi(\rho)\,.
\end{align}
Choosing the unitary matrix in Eq.~\eqref{SABC} to be diagonal, we find that the following boundary conditions are possible:
\begin{align}\label{AppD-BC}
    \Psi\left(\pm\pi/2\right) = \pm\alpha \Psi'\left(\pm\pi/2\right)\,,
\end{align}
where we choose $\alpha >0$. Two independent solutions to Eq.~\eqref{AppD-eigenvalue-eq} with $\omega^2 = - \nu^2 < 0$ are
\begin{subequations}
\begin{align}
    \Psi_\nu^{(1)}(\rho) & = \cosh(\nu\rho)\,,\\
    \Psi_\nu^{(2)}(\rho) & = \sinh(\nu\rho)\,.
\end{align}
\end{subequations}
The functions $\Psi_\nu^{(1)}$ and $\Psi_\nu^{(2)}$ satisfy the boundary conditions~\eqref{AppD-BC} if
\begin{subequations}
\begin{align}
\coth\left(\tfrac{\nu\pi}{2}\right) & = \alpha\nu\,,\label{AppD-BC-eq1}\\
\tanh\left(\tfrac{\nu\pi}{2}\right) & = \alpha\nu\,,
\label{AppD-BC-eq2}
\end{align}
\end{subequations}
respectively.  Equation~\eqref{AppD-BC-eq1} has a solution for all $\alpha >0$ whereas Eq.~\eqref{AppD-BC-eq2} has a solution if $0 < \alpha < \pi/2$.  It is interesting that in the limit $\alpha\to 0$ (the Dirichlet limit) we have $\nu \to \infty$ and hence $\omega^2 \to -\infty$.

%

\end{document}